\documentclass[aps,prr,notitlepage,%
superscriptaddress,twocolumn]{revtex4-1}

\usepackage{natbib}
\usepackage{graphicx}
\usepackage{amssymb}
\usepackage{amsmath}
\usepackage{ifthen}
\usepackage{braket}
\usepackage{xcolor}
\usepackage{bm}
\usepackage{bbm}
\usepackage{comment}
\usepackage{siunitx}
\usepackage{float}
\usepackage{dcolumn}
\usepackage{cancel}

\newcolumntype{d}[1]{D{.}{.}{#1}}
\newcolumntype{.}{D{x}{}{9}}
\newcolumntype{,}{D{x}{}{5}}
\newcolumntype{;}{D{x}{}{8}}

\usepackage{fancyvrb}

\definecolor{garrosgreen}{rgb}{0.1, 0.6, 0.1}
\definecolor{dartmouthgreen}{rgb}{0.05, 0.5, 0.06}
\definecolor{duelferred}{rgb}{0.7, 0.2, 0.1}
\definecolor{cambridgeblue}{rgb}{0.1, 0.3, 1.0}
\definecolor{oxfordblue}{rgb}{0.05, 0.2, 0.7}

\newcommand{\genspin}{S}

\newcommand{\NR}{\mathrm{NR}}

\newcommand{\SO}{\mathrm{SO}}
\newcommand{\SSPIN}{\mathrm{FSS}}

\newcommand{\dd}{\mathrm{d}}
\newcommand{\ii}{\mathrm{i}}
\newcommand{\ee}{\mathrm{e}}

\newcommand{\Dt}{\mathrm{Dt}}

\newcommand{\DT}{{\rm Dt}}

\newcommand{\rmR}{\mathrm{R}}
\newcommand{\Li}{\mathrm{Li}}

\newcommand{\tildeg}{\tilde{g}}

\renewcommand{\Re}{\mathrm{Re}\,}
\renewcommand{\Im}{\mathrm{Im}\,}

\usepackage{xcolor}
\usepackage{bbm}

\definecolor{light}{gray}{0.90}
\definecolor{darker}{gray}{0.50}
\definecolor{dark}{gray}{0.30}

\begin{document}

\title{Bound Deuteron--Antideuteron System (Deuteronium):\\
Leading Radiative and Internal--Structure Corrections to Bound-State Energies}

\author{Gregory S. Adkins}
\affiliation{Department of Physics and Astronomy, Franklin \& Marshall College,
Lancaster, Pennsylvania 17604, USA}

\author{Ulrich D. Jentschura}
\affiliation{Department of Physics and LAMOR, Missouri University of Science and
Technology, Rolla, Missouri 65409, USA}

\begin{abstract}
We evaluate the energy levels of the 
deuteronium bound system, which consists
of a deuteron and an antideuteron,
with a special emphasis on states with
nonvanishing orbital angular momenta.
The excited atomic bound states of deuteronium
constitute probes for the 
understanding of higher-order 
quantum electrodynamic corrections for 
spin-1 particles in a bound system where the typical
field strength of the binding Coulomb field (at a distance 
of the generalized Bohr radius) exceeds Schwinger's
critical field strength.
For states with nonvanishing angular momenta,
effects due to the internal structure
of the deuteron and virtual annihilation
contributions are highly suppressed.
Relevant transitions are found to be in a
frequency range accessible by standard laser spectroscopic
techniques.  We evaluate the leading 
and next-to-leading energy corrections of orders $\alpha^3 m_d$ and 
$\alpha^4 m_d$, where $\alpha$ is the fine-structure
constant and $m_d$ is the deuteron mass,
and also investigate internal-structure corrections:
hadronic vacuum polarization, 
finite-size effects, 
and strong-interaction corrections.
\end{abstract}

\maketitle

%
%
\section{Introduction}

We investigate the 
suitability of deuteronium (the bound system
of a deuteron and its antiparticle,
the antideuteron) for tests of quantum-electrodynamic
interactions involving charged spin-1 
particles.
Deuteronium constitutes an analogue of protonium,
the bound system of proton and 
antiproton~\cite{Ba1989,KlBaMaRi2002},
but is distinct in the particle contents,
adding a neutron and its antiparticle
in the constituent orbiting 
particles. One of the primary points of 
interest for deuteronium is that 
it would allow for the study of higher-order 
effects in spin-1 bound systems,
where the general form of the 
Breit Hamiltonian for particles of 
arbitrary spin~\cite{Pa2007,Pa2008,ZaPa2010}
could be subjected to a thorough test,
and quantum electrodynamic (QED) 
vacuum polarization effects could
be tested in extreme fields.
Specifically, the Coulomb field strength
felt by a constituent particle at the (generalized) Bohr
radius $a_0$ of the bound deuteron-antideuteron 
system (see Table~\ref{table1}) is
$E_C = 1.7320 \times 10^{18} \, \frac{\mathrm{V}}{\mathrm{m}}$
and thus exceeds the Schwinger critical
field strength $E_{\mathrm{cr}} = 
1.3233 \times 10^{18} \, \frac{\mathrm{V}}{\mathrm{m}}$
(for a recent overview of the latter concept,
see Chap.~18 of Ref.~\cite{JeAd2022book}).

The derivation of the spin-1 Breit 
Hamiltonian~\cite{ZaPa2010}
is a nontrivial exercise, which was 
completed by the physics 
community only after combining the results
of independent contributions reported over
the course of a number of decades.
Specifically, after the initial formulation of the relativistic 
spin-1 wave equation~\cite{Pr1936a,Pr1936b}, one
found a suitable conversion of the equation 
to a first-order differential equation, 
to arrive at the Duffin--Kemmer formalism~\cite{Du1938,Ke1939},
which transforms the Proca equation~\cite{Pr1936a,Pr1936b}
into a Dirac-like form.
One then needs to identify a suitable 
wave function for the spin-1 system~\cite{CoSc1940,YoBl1963},
and perform a rather sophisticated 
Foldy--Wouthuysen transformation 
in order to derive the spin-1 Hamiltonian,
first, for a single particle,
and then, for two interacting spin-1 
particles~\cite{Pa2007,Pa2008,ZaPa2010}.

It is appropriate to include a few remarks regarding the 
availability of antideuterons for the production
of the deuteronium bound system.
Antideuterons were initially produced in proton-beryllium collisions many years 
ago (see Refs.~\cite{DoEtAl1965,MaEtAl1965}). Very promising 
approaches to antideuteron production at the AD/ELENA facility~\cite{Ca2024}
and at the GBAR beamlines~\cite{BlOhCr2025}
have recently been explored. One possible 
pathway toward a successful experiment
would involve radiative capture of 
slowed-down antideuteron projectiles in heavy water,
or inside an intense atomic beam containing
deuterium atoms. The captured antideuteron would cascade 
down a chain of radiative decays 
along circular states, lowering the principal 
quantum number in the transitions~\cite{NeEtAl2007}.

In order to fix ideas and provide a useful
reference for the reader, we provide numerical 
values for a number of physical constants
describing properties of the deuteron,
and of deuteronium, in Table~\ref{table1}.
We use natural units with $\hbar = c = \epsilon_0 = 1$ throughout 
this paper, with two exceptions:
one, we restore Syst\`{e}me International (SI mksA) 
units for definiteness in Table~\ref{table1},
and two, in the discussion surrounding 
Eqs.~\eqref{tilderd}, \eqref{tilde_alphaE}, and \eqref{tilde_tauE}.  We adopt
a timelike (``West--Coast'') 
space-time metric, with $g_{00}=+1$, $g_{i j}=-\delta_{i j}$
(where Latin indices are spatial, with $i,j \in \{ 1,2,3 \}$).

The scaling the leading radiative and relativistic 
corrections in deuteronium (with the fine-structure
constant $\alpha$ and the deuteron mass $m_d$) 
is given in Table~\ref{table2}.
The leading one-loop vacuum-polarization correction
(due to the ``Uehling potential'' \cite{Ue1935})
is of order $\alpha^3 m_d$.
This is also the leading correction that leads
to the Lamb shifts in deuteronium.
The (nonradiative)
spin-1--spin-1 Breit interaction
enters at order $\alpha^4 m_d$.
Among the radiative effects of
order $\alpha^4 m_d$, we find
both reducible (loop-by-loop) 
and irreducible~\cite{KaSa1955,LaJe2024}
two-loop vacuum polarization shifts,
and the iterated one-loop electronic vacuum
polarization (eVP) effect in
second-order perturbation theory.
Here, we aim to calculate all corrections to the 
spectrum up to and including the 
next-to-leading order, {\em i.e.},
up to the order $\alpha^4 \, m_d$,
as well as selected corrections 
connected with the internal
structure of the constituent particles.
The hadronic vacuum-polarization
(VP) effect is included in this
list in order to answer a potential question
regarding the importance of virtual 
hadrons in the VP loop, in view
of a small Bohr radius for deuteronium (Dt).
Furthermore, it is indicated to address
other internal-structure effects, such as 
the energy shift due to the deuteron polarizability,
as well as finite-size effects 
and strong-interaction corrections.

\begin{figure}[t!]
\begin{center}
\begin{minipage}{0.99\linewidth}
\begin{center}
\includegraphics[width=0.91\linewidth]{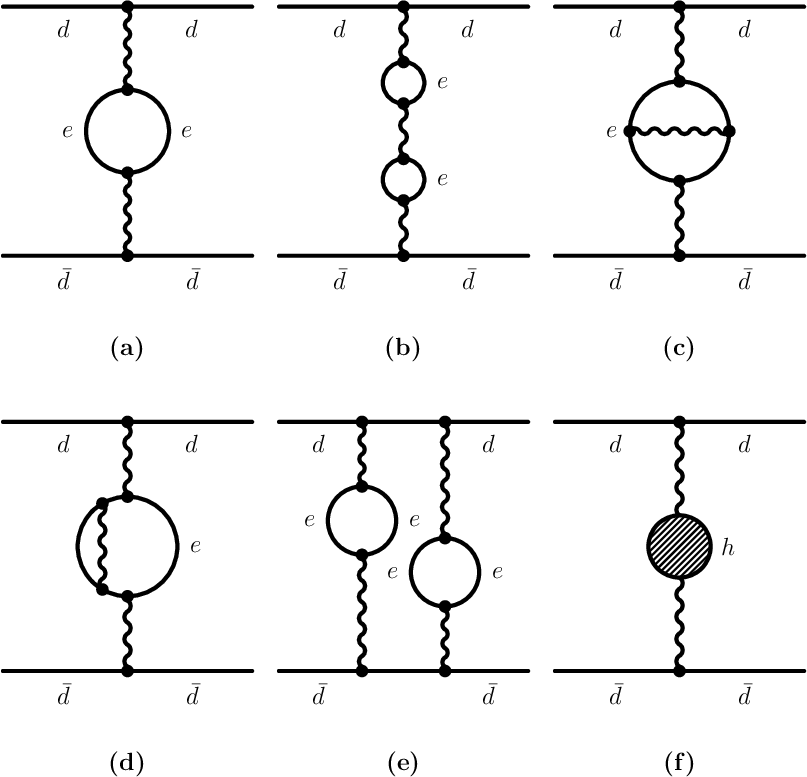}
\end{center}
\caption{\label{fig1} Feynman diagrams contributing to the leading radiative
corrections in deuteronium.  Light fermions (virtual electrons and positrons)
are denoted as ``$e$'', and the hadronic loop is denoted by the letter ``$h$''.
Panel~(a) represents the leading-order eVP correction, (b) is the loop-by-loop
reducible contribution, (c) and (d) are the two independent contributions to
the two-loop irreducible correction, (e) represents the second-order
perturbation theory contribution of two one-loop potentials, and (f) represents
the hadronic vacuum polarization contribution.}
\end{minipage}
\end{center}
\end{figure}

Relevant Feynman diagrams for the dominant 
radiative corrections in deuteronium 
are depicted in Fig.~\ref{fig1};
these are due to vacuum-polarization effects.
In this work, we put special emphasis
on the $2P$, $3P$, $3D$, $4P$, $4D$, and $4F$
states of deuteronium, including their 
spin-dependent (hyper)fine-structure sublevels,
because states with nonvanishing orbital
angular momenta are much less influenced
than $S$ states by strong-interaction 
effects~\cite{DeGoBaTh1954,BedH1955,By1957,Tr1961,KlBaMaRi2002,AdJe2025trueman}.
For $S$ states,
these latter effects are large and challenging to analyze
[see Ref.~\cite{AdJe2025trueman} and Eq.~\eqref{trueman}].

We organize the paper as follows.  In Sec.~\ref{sec2}, we discuss
a number of important properties of the deuteronium system.  
The Breit Hamiltonian is treated in Sec.~\ref{sec3}.
Vacuum polarization corrections are 
discussed in Sec.~\ref{sec4}.
Effects related to the strong interaction and to the internal structure
of the constituent particles are 
discussed in Sec.~\ref{sec5}.
Finally, the allowed dipole
transitions in the system are discussed in Sec.~\ref{sec6}.
Relevant formulas from the angular-momentum 
algebra are discussed in Appendix~\ref{appa}.

\renewcommand{\arraystretch}{1.8}
\begin{table}[t!]
\caption{\label{table1} Some useful parameters
are given for deuteronium (Dt).
We use the CODATA
recommended values of the fundamental constants~\cite{TiMoNeTa2021,MoNeTaTi2025}.
The fine-structure constant is given by $1/\alpha = 137.035\,999\,177(21)$.
For the quadrupole moment, we use the result from 
Ref.~\cite{PuKoPa2020}. The quadrupole moment of the deuteron
is divided by the elementary charge,
following commonly adopted 
conventions (see Refs.~\cite{KeRaRaZa1939,KeRaRaZa1940,PuKoPa2020})}
\begin{tabular}{l@{\hspace*{.4cm}}l}
\hline
\hline
Deuteron mass & $m_d c^2 = 1875.612\,945\,00(58) \, {\rm MeV}$ \\
Reduced mass &
$m_r c^2= \frac12 m_d c^2 = 937.806\,472\,50(29) \, {\rm MeV}$ \\
Dt Bohr radius &
$\displaystyle{a_0 = \frac{\hbar}{\alpha m_r c} = 28.834\,200\,570(10) \, {\rm fm}}$ \\
Dt Hartree &
$E_h = \alpha^2 \, m_r \, c^2 
= 49.939\,464\,871(22) \, {\rm keV}$ \\
Dt Rydberg &
$E_0 = \frac14 \, \alpha^2 \, m_d \, c^2
= 24.969\,732\,435(11) \, {\rm keV}$ \\
NR Spectrum &
$\displaystyle{E_{\rm NR} = 
- \frac{E_0}{n^2} = - \frac12 \, \frac{E_h}{n^2}
= -\frac{\alpha^2 \, m_d}{4 n^2}}$ \\
Dt $\beta$ Parameter & 
$\displaystyle{\beta_\DT = \frac{m_e}{\alpha m_r}
= 0.037\,334\,596\,12(12) }$ \\
Deuteron $g$ factor &
$g_d = 0.857\,438\,2335(22)$ \\
Scaled $g$ factor &
$\tilde g_d = (m_d/m_p) g_d = 1.714\,025\,4606(45)$ \\
Deuteron radius & 
$r_d = 2.12778(27)  \, {\rm fm}$ \\
Quad.~moment & 
$Q_{d} = 0.285\,699(24) \, {\rm fm}^2$  \\
\hline
\hline
\end{tabular}
\end{table}

\renewcommand{\arraystretch}{1.3}
\begin{table}[t!]
\begin{center}
\caption{\label{table2}
We list the order-of-magnitude (or physical origin) of several 
energy corrections in deuteronium.
We evaluate all contributions
up to and including those of order
$\alpha^4 \, m_d$, and also 
some conceptually interesting
effects that are linked to the strong interaction and to the 
internal structure of the constituent particles of Dt.  
}
\begin{tabular}{l@{\hspace*{1cm}}l}
\hline
\hline
Effect & Order \\
\hline
One-loop eVP & $\alpha^3 m_d$ \\
\hline
Breit Hamiltonian  & $\alpha^4 m_d$ \\
\hline
Reducible two-loop eVP       & $\alpha^4 m_d$ \\
Irreducible two-loop eVP   & $\alpha^4 m_d$ \\
Second--order eVP       & $\alpha^4 m_d$ \\
\hline
Hadronic VP             & photon structure \\
Finite--size            & deuteron structure \\
Deuteron polarizability & deuteron structure \\
Strong interaction & strong interaction \\
\hline
\hline
\end{tabular}
\end{center}
\end{table}

%
%
\section{Basic Properties}
\label{sec2}

In the center-of-mass frame, the nonrelativistic Hamiltonian
of deuteronium reads as follows,
\begin{equation}
H_{\rm NR} = \frac{\vec p^{\,2}}{m_d} - \frac{\alpha}{r} \,,
\end{equation}
where $\vec p$\, is the relative momentum operator in that frame.
Convenient quantum numbers for deuteronium 
are the principal quantum number $n$,
the orbital angular momentum quantum number $L$,
the total spin quantum number $S$ 
describing the magnitude of the total spin $\vec S = \vec S_+ + \vec S_-$
(where $\vec S_+$ and $\vec S_-$ are the 
spin operators for the deuteron and its antiparticle),
the total angular momentum quantum number $J$,
and its projection $J_z$.
We denote the states as
\begin{equation}
| n {}^{2 S + 1} \! L_J (J_z) \rangle =
| n L S J J_z \rangle \, ,
\end{equation}
making use of the spectroscopic notation $n {}^{2 S + 1}L_J $
(in which states with $L=0$
are $S$ states, states with $L=1$ are $P$ states, etc.).
The nonrelativistic wave functions are
\begin{equation}
\label{psi}
\psi_{n LSJ J_z}(\vec r \, ) = R_{nL}(r) \,
\Xi^{L S}_{J J_z}(\theta, \varphi) \,,
\end{equation}
where $R_{nL}(r)$ is a Schr\"{o}dinger--Coulomb radial 
eigenfunction of a hydrogenlike ion
with nuclear charge number $Z=1$ and
reduced mass $m_r = m_d/2$. 
Specifically (see, {\it e.g.}, Chap.~4 of Ref.~\cite{JeAd2022book}) one has
\begin{equation} \label{radial_wf}
R_{n L}(r) = \left ( \frac{4 (n-L-1)!}{a_0^3 n^4 (n+L)!} \right )^{1/2}
\rho^L \ee^{-\rho/2} L_{n-L-1}^{2L+1}(\rho) \,,
\end{equation}
where $a_0 = 1/(\alpha m_r) = 2/(\alpha m_d)$
is the (generalized) Bohr radius, and
$\rho = 2 r/(a_0 n)$.
The angular part of the wave function
for the state $| n {}^{2 S + 1} \! L_J (J_z) \rangle$ is
\begin{equation}
\label{defchi1}
\Xi^{L S}_{J J_z}(\theta, \varphi) =
\sum_{M_L M_S} C^{J J_z}_{L M_L ; S M_S} \, 
Y_{L M_L}(\theta, \varphi) \, \chi_{S M_S} \, ,
\end{equation}
where $\chi_{S M_S}$ is the deuteronium total spin state
\begin{equation}
\label{defchi2}
\chi_{S M_S}
=  \sum_{m_+ m_-}
C^{S M_S}_{1 m_+ ;1 m_-} \;
\chi^{(+)}_{m_+} \;
\chi^{(-)}_{m_-} \,.
\end{equation}
Here, the $\chi^{(\pm)}_{m_\pm}$ are the eigenspinors
of the spin operators of the deuteron and 
the antideuteron. A convenient representation of the 
spin-1 matrices is 
$(\genspin_\pm^i)^{j k} = -\ii \epsilon^{i j k}$.
These matrices act on the deuteron~($+$) and 
antideuteron~($-$) spinors. They fulfill the relations
$(\genspin_\pm^z) \, \chi^{(\pm)}_{m_\pm} = 
m_\pm \chi^{(\pm)}_{m_\pm}$, where $m_\pm \in \{-1,0,+1\}$.
In Eqs.~\eqref{defchi1} and \eqref{defchi2}, the 
Clebsch--Gordan coefficients 
$C^{J M_J}_{j_1 m_1 ; j_2 m_2}$ are used
in the notation of Chap.~6 of Ref.~\cite{JeAd2022book}.

The lowest-order quantum mechanical eigenproblem
is solved by 
\begin{equation}
H_{\rm NR} \, 
\psi_{n LSJ J_z}(\vec r \, ) = 
E_{\rm NR} \,
\psi_{n LSJ J_z}(\vec r \, ) \,.
\end{equation}
It gives rise to the nonrelativistic 
bound-state energy levels which depend
only on the principal quantum number~$n$,
\begin{equation}
E_{\rm NR} = -\frac{\alpha^2 m_d}{4 n^2} 
= -\frac{E_0}{n^2} 
\approx - \frac{24969.732 \,{\rm eV}}{n^2}  \,,
\end{equation}
where the Dt Rydberg constant $E_0$
is defined in Table~\ref{table1}.  One should 
take into consideration that, strictly speaking,
the spin-dependent Breit Hamiltonian 
(see the following section) is not completely diagonal in the
$\psi_{n L S J J_z}$ basis, so that certain
mixings between states with ``good'' quantum numbers
occur in higher orders. Analogous effects occur
in hydrogenlike systems, where the effect is
otherwise known as the fine-structure hyperfine-structure
mixing (for a particularly instructive discussion,
in the context of the mixing of the fine and the hyperfine 
structure for muonic deuterium, 
see Sec.~6 of Ref.~\cite{KrEtAl2016deuteronium}).

The presence of the mixing terms indicates that a careful consideration of the
conserved quantum numbers of deuteronium is indicated. The total angular
momentum $J$ and its $z$-component $J_z$ are, of course, conserved.  Parity and
charge parity are conserved in QED.  The parity of the antideuteron must be
positive, since it is determined by the intrinsic parities of its constituents
(both $-1$) and the parity of the antideuteron quantum state, which is positive
since it has the same structure as for the deuteron, being a combination of $S$
and $D$ waves.  Then, the parity of a deuteronium state is determined by the
parities of its constituents (both +1) and that of its quantum state $(-1)^L$,
for a total parity $P_{\rm Dt} = (-1)^L$.  So the oddness or evenness of $L$ is
conserved.  Deuteronium, being formed of a particle and its antiparticle, can
also exist in an eigenstate of charge parity.  We determine the charge
conjugation quantum number $C_{\rm Dt}$ by considering what happens to a
deuteronium state when its constituents are interchanged.  Since 
deuterons are bosons, 
the effect of this interchange is a factor $+1$ for the wave function.
A complete interchange involves the exchange of spatial states [with a factor
of $(-1)^L$], spin states [factor $(-1)^{S}$], and charge states (factor
$C_{\rm Dt}$).  We find that $+1 = (-1)^L \, (-1)^{S} \, C_{\rm Dt}$, so that
$C_{\rm Dt} = (-1)^{L+S}$.  In all, we have
\begin{equation}
P_{\rm Dt} = (-1)^L \, , \; C_{\rm Dt} = (-1)^{L+S} \,, \; 
P_{\rm Dt} C_{\rm Dt} = (-1)^S \, .
\end{equation}
So the oddness or evenness of $S$ is also conserved.

Good states of deuteronium are specified by specific values of $n$, $J$, and
$J_z$.  In addition, they are constructed of states with either even or odd
values of $L$, and either even or odd values of $S$.  One might thus assume
that the Breit Hamiltonian could mix states with different values of $L$
provided both states involved in the mixing are even, or, both are odd.
However, it is known that the spin-1/2--spin-1/2 Breit Hamiltonian does not mix
states with different values of $L$.  Since the spin-1--spin-1 Breit
Hamiltonian is built of operators with the same types of spatial dependence, it
also does not mix states with different values of $L$.  
So, $L$ can be considered to be a good quantum number (to all orders).

The caveat for the total spin is that the Breit Hamiltonian, indeed,
mixes spin states, so deuteronium states with even spin 
($S=0$ and $S=2$) can be mixed. Odd-spin states have $S=1$. 
It means that, strictly speaking, $S$ is only a good quantum
number up to order $\alpha^4 m_d$, when the Breit Hamiltonian 
starts mixing the spin states. However, the same could
be said about the principal quantum number $n$,
since the Uehling potential mixes states with different values of $n$.
We here deal with the Breit Hamiltonian (and the Uehling potential)
perturbatively, including
contributions of first, second, and (eventually) third-order perturbation
theory.

It is instructive to explore the multiplicity
and (hyper)fine-structure of the bound states
of deuteronium for given principal quantum number $n$
and orbital angular momentum $L$.
For $S$ states ($L=0$), there are three states with $S=J$ equal to $0$, $1$,
or $2$. For $P$ states ($L=1$), when $S=0$, 
there is only one possible value of $J$, namely, $J=1$. When $S=1$,
$J$ can take any of the three possible values $0$, $1$, and $2$.
When $S=2$, there are also three possible values of 
$1 \leq J \leq 3$.  In all,
there are seven $P$ states. 
In view of the above mentioned symmetry considerations,
the only ones that can mix have $S=0$, $J=1$ and
$S=2$, $J=1$.  In general, the states that mix have $S=0$ or $S=2$ with $L=J$.
For $D$ states ($L=2$), when $S=0$, the only 
possible value of $J$ is
$J=2$.  When $S=1$, there are three possible values of $J$,
namely, $1$, $2$, and $3$.  When
$S=2$, $J$ can take any value in the set
$J \in \{ 0,1,2,3,4 \}$.
In all, there
are nine $D$ states.  The only ones that can mix have $S=0$, $J=2$ and $S=2$,
$J=2$.  For states with $L > 2$, there are also nine states, and the only ones
that can mix have $S=0$, $J=L$ and $S=2$, $J=L$.  The states are explicitly
displayed in Table~\ref{table3}.  We label the states using spectroscopic
notation as $n^1 \!L_J$ for $S=0$, $n^3 \!L_J$ for $S=1$, and $n^5 \!L_J$ for $S=2$.
The two that mix are $n^1 \!L_L$ and $n^5 \!L_L$.  The states that diagonalize the
Breit Hamiltonian are labeled $n^- \!L_L$ and $n^+ \! L_L$, where the first has
the smaller energy and the second the larger energy.
For absolute clarity, we reemphasize that 
the superscript $\pm$, in this notation,
denotes the superposition of $S=0$ and $S=2$ states,
and has nothing to do with the parity of the bound state.

The leading corrections to the energies of deuteronium are due to the
strong interaction and to electron vacuum polarization.  The strong
interaction shifts are most pronounced in $S$ states, since the ratio
of the range of the strong force ($\approx 1\,{\rm fm}$) to the 
(generalized) Bohr radius
of deuteronium ($a_0 \approx 28.8 \, {\rm fm}$) is relatively small.
Strong interaction shifts are noticeable in $P$ states and negligible
for $L \ge 2$.  Electron vacuum polarization (the Uehling correction \cite{Ue1935}) 
is by far the leading correction for states with $L \ge 1$, being of order
$\alpha^4 \, m_d$.  The corrections
due to electron vacuum polarization are spin-independent, depending
only on $n$ and $L$, and can be said to give rise to 
Lamb shifts such as for $2P$--$2S$.  
Spin-dependent shifts commence at order $\alpha^4 m_d$ with the Breit correction,
as described in the following section.

%
%
\section{Breit Hamiltonian}
\label{sec3}

In order to derive the spin-dependent corrections
of order $\alpha^4 m_d$ to the spectrum,
we generalize the Breit Hamiltonian
to the interaction
of two spin-1 particles.
We use the results communicated in
Eq.~(3) of Ref.~\cite{Pa2007} and
Eq.~(17) of Ref.~\cite{Pa2008},
and Eq.~(12.100) of Ref.~\cite{JeAd2022book}.
An alternative derivation,
which starts from the
single-particle Hamiltonian for general
spin [see  Eqs.~(1)---(4) of Ref.~\cite{ZaPa2010}]
has recently been outlined in Ref.~\cite{AdJe2025RReVP}.
The latter derivation starts from a
single-particle spin-1 Hamiltonian,
identifies the corresponding Feynman
rules of Nonrelativistic Quantum 
Electrodynamics (NRQED), and leads to the
interaction Hamiltonian based on the
single-photon interaction kernels
identified in Ref.~\cite{AdJe2025RReVP}.

The Breit Hamiltonian $H_{\rm BR}$ gives
corrections of order $\alpha^4 \, m_d$ 
to the spectrum. Based on the considerations reported in
Refs.~\cite{AdJe2025RReVP,AdJe2025prl},
we conveniently write the Breit 
Hamiltonian $H_{\rm BR}$ for an interacting system of two spin-1 particles 
as follows,
\begin{align}
H_{\rm BR} =& \; H_{\rm K} + H_{\rm M} + H_{\SO} + H_{\SSPIN} + H_{\rm Q} + H_D \,.
\end{align}
This Hamiltonian is the sum of the kinetic term $H_{\rm K}$,
which describes the relativistic corrections 
to the kinetic energies of the constituent particles,
the magnetic term $H_{\rm M}$, which describes
the exchange of a transverse photon with convection vertices at each end,
the spin-orbit term $H_{\rm SO}$,
the Fermi spin-spin term $H_{\SSPIN}$,
the quadrupole term $H_{\rm Q}$ which is due to the intrinsic 
quadrupole moment of the deuteron (and anti-deuteron),
and the Dirac delta term $H_D$,
which describes the 
deuteron finite-size effect, within the Dirac delta 
approximation (a more thorough treatment is presented
in Sec.~\ref{sec5B}).
For convenience, we list them as follows,
\begin{subequations}
\label{Hcontrib}
\begin{align}
H_{\rm K} =& \; - \frac{\vec p^{\,4}}{4 m_d^3} \,,
\quad
H_{\rm M} = - \frac{\alpha}{2 m_d^2} \,
p^i \, \left( \frac{\delta^{ij} + \hat x^i \, \hat x^j}{r} 
\right) \, p^j \,,
\\
H_{\rm SO} =& \;
\frac{\alpha \, (2 \tilde g_d - 1 )}{2 \, m_d^2 r^3} \,
(\vec \genspin_+ + \vec \genspin_-) \cdot \vec L \,,
\\
H_{\SSPIN} =& \; \frac{2 \pi \alpha \, \tilde g_d^2}{3 m_d^2} \,
\vec \genspin_+ \cdot \vec \genspin_- \, \delta^{(3)}(\vec r \, )  
\nonumber\\
& \; 
+ \frac{3 \alpha \, \tilde g_d^2}{4 m_d^2 \, r^3} \, 
(\genspin_+^i \genspin_-^j)^{(2)} \, (\hat x^i \, \hat x^j)^{(2)} 
\,,
\\
H_{\rm Q} \!  =& \!
- \! \frac{3 \, \alpha \, \tilde Q_{d} }{2 m_d^2 r^3} \, 
\big [ (\genspin_+^i \genspin_+^j)^{(2)} \! + 
(\genspin_-^i \genspin_-^j)^{(2)} \big ] \, (\hat x^i \, \hat x^j)^{(2)} ,
\\
\label{Hdelta}
H_{\rm D} =& \;
\frac{4 \pi \alpha}{3} \frac{\tilde r_d^2}{m_d^2} \, \delta^{(3)}(\vec r \, )  \,.
\end{align}
\end{subequations}
We recall that Latin superscripts denote spatial components ($i,j = 1,2,3$).
The unit vector is $\hat x = \vec r/| \vec r \, |$.
Also, $(\genspin^i_\pm \genspin^j_\pm)^{(2)}$
is the quadrupole component of a spin-spin tensor,
where the quadrupole component of the tensor product 
of two vectors $\vec v $ and $\vec w$ 
(which do not necessarily commute)
is $( v^i \, w^j )^{(2)} = \tfrac12 ( v^i \, w^j + v^j w^i ) - 
\tfrac13 \delta^{ij} \, \vec v \cdot \vec w$.
Data for the deuteron radius, its electric quadrupole moment,
and its $g$ factor (and scaled $g$ factor) can
be found in Table~\ref{table1}.  
We note that the dimensionless values of 
the deuteron radius, $r_d$, and 
the deuteron quadrupole moment, $Q_{d}$, are
(see Refs.~\cite{TiMoNeTa2021,MoNeTaTi2025,PuKoPa2020})
\begin{equation} \label{tilderd}
\tilde r_d \equiv m_d \, r_d = \frac{m_d \, c}{\hbar} r_d = 20.2248(26)
\end{equation}
and
\begin{equation} \label{tildeQd}
\tilde Q_d = m_d^2 \, Q_{d} = 
\left( \frac{m_d \, c}{\hbar} \right)^2 \, Q_{d} = 25.8120(22) \, ,
\end{equation}
leading to relatively large contributions from these quantities.  

The Breit Hamiltonian is derived from 
the one-photon exchange between the constituent
particles of deuteronium (no virtual loops) and is the 
parametrically dominant {\em nonradiative} 
correction to the spectrum of deuteronium (order $\alpha^4 m_d$).
Moreover, the Breit Hamiltonian leads to 
dominant spin-dependent energy corrections in deuteronium.
For comparison, we point out 
that (spin-independent) radiative effects due to vacuum polarization
(with a virtual electron-positron loop)
are parametrically larger than the Breit 
corrections and commence at order $\alpha^3 m_d$. These
affect the spin-independent structure of deuteronium, with dependence
only of the quantum numbers $n$ and $L$, as
discussed in Sec.~\ref{sec4A}. Higher-order electron vacuum
polarization corrections (which are still spin-independent 
effects) occur at order $\alpha^4 m_d$ and are 
discussed in Secs.~\ref{sec4C}---\ref{sec4E}. 
Here, we first treat the spin-dependent energy 
correction $E_{\rm BR}$ which appears
at order $\alpha^4 m_d$ due to the Breit Hamiltonian:
\begin{equation}
E_{\rm BR} = \langle H_{\rm BR} \rangle = 
\langle n LSJ J_z | H_{\rm BR} | n LSJ J_z \rangle \, .
\end{equation}

We obtain the corrections coming from the Breit
Hamiltonian one term at a time.
We start with the kinetic term $H_{\rm K}$.
In order to evaluate the expectation value 
of $\vec p^{\,4}$, one observes that
the momentum operator acts 
on wave functions as follows
(in our short-hand notation, $\psi$
stands for $\psi_{nLSJ J_z}$),
\begin{equation}
\label{action_p2}
\vec p^{\,2} | \psi \rangle = 
m_d \, \left( E_\NR + \frac{\alpha}{r} \right)
| \psi  \rangle \,,
\end{equation}
So, one has the relation (we use a further 
shorthand notation $\langle K \rangle \equiv
\langle \psi | K | \psi \rangle$ for any operator $K$),
\begin{multline}
\label{E_K}
E_{\rm K} = \left< -\frac{\vec p^{\,4}}{4 m_d^3} \right> 
= -\frac{1}{4 m_d} \, 
\left< \left( E_\NR + \frac{\alpha}{r} \right)^2 \right>
\\
= -\frac{1}{4 m_d} \, \left< E_\NR^2 + 2 E_\NR \frac{\alpha}{r} + 
\left( \frac{\alpha}{r} \right)^2 \right>
\\
= \alpha^4 m_d \left \{ \frac{3}{64n^4} - \frac{1}{8n^3 (2L+1)} \right \} \, .
\end{multline}
We have used the following
matrix elements from Eq.~(4.346) of Ref.~\cite{JeAd2022book},
\begin{subequations}
\label{matelems}
\begin{align}
\left< \frac{1}{r} \right> =& \;
\frac{\alpha m_d}{2 n^2} \,,
\\
\left< \frac{1}{r^2} \right>
=& \; \frac{\alpha^2 m_d^2}{2 n^3\,(2\,L+1)}\,.
\end{align}
\end{subequations}
The magnetic term $H_{\rm M}$ is conveniently 
evaluated by using the following identity,
\begin{multline}
H_{\rm M} = - \frac{\alpha}{2 m_d^2} \,
p^i \left( \frac{\delta^{ij}}{r} +
\frac{r^i \, r^j}{r^3} \right) \, p^j
\\
= - \frac{\alpha}{2 m_d^2 r} \,
\left( \vec p^{\,2} +
\frac{1}{r^2} [ \vec r \cdot ( \vec r \cdot \vec p \, ) \vec p \, ] \, 
\right) \,.
\end{multline}
This relation is consistent with Eq.~(84.1) of Ref.~\cite{BeLiPi1982vol4}.
Use of Eq.~\eqref{action_p2} leads to the 
appearance of matrix elements proportional to 
$\langle 1/r^3 \rangle$, which can be 
evaluated using Eq.~(4.346) of Ref.~\cite{JeAd2022book},
\begin{equation}
\label{onebyr3}
\left< \frac{1}{r^3} \right> =
\frac{\alpha^3 \, m_d^3}{4 n^3\,L\,(L+1)\,(2 L+1)}\,,
\end{equation}
which is divergent for $S$ states with $L=0$.
However, a closer inspection reveals that the 
matrix element $E_{\rm M} = \langle H_{\rm M} \rangle$
is convergent for $S$ states, and one obtains
the general formula
\begin{equation}
\label{E_M}
E_{\rm M} = \frac{ \alpha^4 m_d}{8 n^4} 
+ \frac{\alpha^4 m_d}{8 n^3} \delta_{L 0}
- \frac{3 \alpha^4 m_d}{8 n^3 (2L + 1)}  \,.
\end{equation}

The next term is the spin-orbit term $H_{\rm SO}$.
One writes the total spin operator 
(sum for both particles) as 
$\vec S = \vec\genspin_+ + \vec\genspin_-$,
and obtains
\begin{equation}
E_{\SO} = \frac{\alpha (2 \tildeg_d - 1 )}{2 \, m_d^2} \,
\langle \vec L \cdot \vec S \, \rangle
\left \langle \frac{1}{r^3} \right \rangle \,.
\end{equation}
With use of the definition
\begin{equation}
\label{B_LSJ}
B_{LSJ} = \frac12 \big [ J(J+1) - L(L+1) - S(S+1) \big ]
\end{equation}
one has
\begin{equation}
\label{defBLSJ}
\langle \vec L \cdot \vec S \, \rangle = B_{LSJ} \,.
\end{equation}
So,
\begin{equation}
\label{E_SO}
E_{\SO} = \frac{\alpha (2 \tildeg_d - 1 )}{2 \, m_d^2} \,
B_{LSJ} \, \left< \frac{1}{r^3} \right> \,.
\end{equation}
The remaining matrix element $\langle 1/r^3 \rangle$ is evaluated 
using Eq.~\eqref{onebyr3}.
The energy shift $E_{\SO}$ vanishes for 
$S$ states by spherical symmetry (despite the apparent $1/L$ singularity when $L=0$).

The scalar part of the Fermi-spin-spin energy correction,
\begin{equation}
\label{E_FSS_scalar}
E_{\SSPIN,1} = \; 
\frac{2 \pi \alpha \, \tilde g_d^2}{3 m_d^2} \,
\langle \vec \genspin_+ \cdot \vec \genspin_- \; 
\delta^{(3)}(\vec r \, ) \rangle \,,
\end{equation}
is relatively easy to evaluate.
First, one realizes that 
the expectation value of the Dirac delta is
$\left \langle \delta^{(3)}(\vec r \, ) \right \rangle
= \frac{(\alpha m_d)^3}{8 \pi n^3} \, \delta_{L 0}$.
With the help of the identity
\begin{equation}
\langle \vec \genspin_+ \cdot \vec \genspin_- \rangle =
\frac12 \, \big [ S ( S + 1) - 4 \big ] \,,
\end{equation}
one obtains the result
\begin{equation}
\label{delE_FSS1}
E_{\SSPIN,1} = {\tilde g}_d^2 \,
\frac{\alpha^4 \, m_d}{24 \, n^3} \,
\big [ S(S+1)-4 \big ] \, \delta_{L 0} \,.
\end{equation}

For the tensor part of the Fermi spin-spin term,
\begin{equation}
H_{\rm FSS,2} = \frac{3 \alpha \, \tilde g_d^2}{4 m_d^2 \, r^3} \, 
(\genspin_+^i \genspin_-^j)^{(2)} \, (\hat x^i \, \hat x^j)^{(2)} 
\end{equation}
and the quadrupole term
\begin{equation}
H_{Q} \!  = \!
- \! \frac{3 \, \alpha \, \tilde Q_d }{2 m_d^2 r^3} \, 
\big [ (\genspin_+^i \genspin_+^j)^{(2)} \! + 
(\genspin_-^i \genspin_-^j)^{(2)} \big ] \, (\hat x^i \, \hat x^j)^{(2)}  \, ,
\end{equation}
one needs to carry out some 
more angular algebra.  We need to evaluate matrix elements involving
different values of the total spin instead of simply expectation values, 
since these terms are non-diagonal in the spin variable.
We thus consider an ``almost diagonal'' matrix element 
$\big \langle f(r) M \big \rangle_{n L S' S J} \equiv \langle n L S' J J_z | f(r) M | n L S J J_z \rangle$,
where all quantum numbers of the bra and ket states
are equal except, possibly, for different $S$ and $S'$.
The operator $f(r)$ is supposed to have
only a radial dependence, and $M$ is assumed to 
depend only on the elements of the unit vector $\hat x$,
and on the spin variables.
With reference to the radial component
of the wave function [see Eq.~\eqref{psi}],
we can first separate out the radial dependence using
\begin{equation}
\Big \langle f(r) M \Big \rangle_{n L S' S J} =
 \int \dd r \, r^2 \, \big [ R_{nL}(r) \big ]^2 \, f(r)
\; \big \langle M \big \rangle_{L S' S J}
\end{equation}
when $M$ is independent of the radial variable and
\begin{equation}
\big \langle X \big \rangle_{L S' S J} \equiv 
\big \langle L S' J \vert X \vert L S J \big \rangle 
\end{equation}
for any operator $X$.
Employing relations derived in Appendix~\ref{appa},
which are based on angular reduction,
one derives the following angular matrix elements:
\begin{eqnarray} \label{def_DLSJ}
D_{L S' S J} &\equiv& \left \langle 
\big [ (S_+^i \, S_+^j)^{(2)} + (S_-^i \, S_-^j)^{(2)} \big ]
( \hat x^i \hat x^j )^{(2)} \right \rangle_{L S' S J} 
\nonumber \\
&\phantom{=}& \hspace{-0.8cm} = (-1)^{J+S'} \Big [ (-1)^S + (-1)^{S'} \Big ] 
\sqrt{ \frac{10}{3} } (2L+1) \nonumber \\
&\phantom{=}& \hspace{-0.8cm} \times \Pi_{S' S} 
\begin{Bmatrix} J & S' & L  \\ 2 & L & S  \end{Bmatrix} 
\begin{pmatrix} L & 2 & L \\ 0 & 0 & 0 \end{pmatrix} 
\begin{Bmatrix} S & S' & 2  \\ 1 & 1 & 1  \end{Bmatrix} \, ,
\end{eqnarray}
and
\begin{eqnarray} \label{def_CLSJ}
C_{L S' S J} &\equiv& \left \langle  (S_+^i \, S_-^j)^{(2)} 
( \hat x^i \hat x^j )^{(2)} \right \rangle_{L S' S J} \nonumber \\
&=& (-1)^{(S+S')/2} \, \frac{1}{2} (1+\delta_{S' S} ) D_{L S' S J} \, ,
\end{eqnarray}
where $\Pi_{a b \cdots} \equiv \sqrt{(2a+1) (2b+1) \cdots}$.
Standard notation is used for the 
Wigner $3j$ and $6j$ symbols
(see Chap.~6 of Ref.~\cite{JeAd2022book}).
One thus obtains the result
\begin{eqnarray} \label{matrix_FSS2_Q}
\big \langle H_{\rm FSS,2} + H_{\rm Q} \big \rangle_{n L S' S J} \nonumber \\
&\phantom{x}& \hspace{-3.2cm} 
= \frac{3 \alpha}{4 m_d^2} 
\big \{ \tilde g_d^2  C_{L S' S J} - 2 \tilde Q_d D_{L S' S J} \big \} 
\left \langle \frac{1}{r^3} \right \rangle \nonumber \\
&\phantom{x}& \hspace{-3.2cm} = 
\frac{m_d \alpha^4}{n^3} \times 
\frac{ 3 \big ( \tilde g_d^2  C_{L S' S J} - 
2 \tilde Q_d D_{L S' S J} \big ) }{16L (L+1)(2L+1)} \, .
\end{eqnarray}
Both $C_{L S' S J}$ and $D_{L S' S J}$ vanish when $L=0$ by
spherical symmetry, and this zero in the numerator takes precedence over the
apparent singularity caused by $L$ in the denominator.

Finally, for the Dirac delta contribution, one has
\begin{equation}
\label{E_D}
E_D =
\frac{4 \pi \alpha}{3 m_d^2} \tilde r_d^2 \, 
\langle \delta^{(3)}(\vec r \, ) \rangle =
\frac{\alpha^4 m_d}{6 n^3} \, \tilde r_d^2 \, \delta_{L0} \,.
\end{equation}

Let us fix the quantum number $n$, $L$, $J$,
and $J_z$, and consider the matrix 
elements of the Breit Hamiltonian
in the basis of the total spin quantum
numbers $S$ and $S'$, which takes values
equal to $0$, $1$ and $2$. In the basis of the 
$S$ and $S'$ quantum numbers, one thus 
obtains a three-by-three matrix.
According to the above mentioned symmetry
considerations, the resulting matrix is
diagonal except for the case $L = J$, 
where one encounters off-diagonal elements
for $S' = 2$, $S = 0$ and $S'= 0$, $S = 2$.
Summing up the results given 
in Eqs.~\eqref{E_K}, \eqref{E_M}, \eqref{E_SO}, 
\eqref{delE_FSS1}, \eqref{matrix_FSS2_Q} and~\eqref{E_D}, 
one obtains 
\begin{eqnarray}
\left \langle H_{\rm BR} \right \rangle_{n L S' S J J_z}
&=& \left \langle n L S' J J_z \vert H_{\rm Br} 
\vert n L S J J_z \right \rangle \nonumber \\
&\equiv& \alpha^4 m_d \, M^{n L J}_{S' S} \,.
\end{eqnarray}
The results are,
of course, independent of $J_z$.  The explicit
expression for $M^{n L J}_{S' S}$ is
\begin{eqnarray}
\label{MBR}
M^{n L J}_{S' S}  &=&
\left ( \frac{11}{ 64 n^4 } - \frac{1}{ 2 n^3 (2L+1)} \right ) 
\delta_{S' S} \nonumber \\
&+& \frac{\delta_{L\,0}}{8 n^3} \left ( 1 + \frac43
\tilde r_d^2 + \frac{\tilde g_d^2}{3} \big [ S(S+1)\!-\!4 \big ] \right ) 
\delta_{S' S} \nonumber \\
&+& \frac{ (2 \tilde g_d - 1) \, 
B_{LSJ} }{8 n^3 L (L+1) (2L+1) } \delta_{S' S} \nonumber \\
&+& \frac{3 \big ( \tilde g_d^2 
C_{L S' S J} - 2 \tilde Q_d D_{L S' S J} \big ) }{16 n^3 L (L+1)(2L+1)} \,.
\end{eqnarray}
The angular structures $B_{LSJ}$, $C_{LS'SJ}$, and $D_{L S' S J}$ are
defined in Eqs.~\eqref{B_LSJ},~\eqref{def_CLSJ} and~\eqref{def_DLSJ}.  

For most values of $L$ and $J$, the $M^{n L J}_{S' S}$ matrices are 
diagonal, but when $L=J$, the 
submatrix spanned by the spin quantum numbers
$S',S \in \{0,2\}$ is not diagonal.  
For example, for $n=2$, $L=J=1$, the $M^{2,1,1}_{S' S}$ matrix 
is a $(3 \times 3)$-matrix in the basis with $S',S \in \{0,1,2\}$
and reads as follows,
\begin{equation}
\label{M211}
M^{2,1,1} = \begin{pmatrix} -0.01009 & 0 & 0.12367 \\ 
0 & 0.05463 & 0 \\ 0.12367 & 0 & 0.05969 \end{pmatrix} \,.
\end{equation}
The element in the center of the matrix corresponds to the 
case $S = S' = 1$ and does not couple to the 
states with other values of $S$. The eigenvalues of the 
matrix are $-0.10369$, $0.05463$, and
$0.15329$.  The states $n^-L_L$ and
$n^+L_L$ are constructed using the $S=0$ and $S=2$ states $n^1L_L$ 
and $n^5L_L$ to diagonalize the Breit Hamiltonian, with $n^-\!L_L$ ($n^+\!L_L$) being
associated with the lower (higher) energy eigenvalue.
Numerical results for Breit energies are presented in Table~\ref{table3}.

\renewcommand{\arraystretch}{1.0}
\begin{table}[t!]
\begin{center}
\caption{\label{table3} Relativistic corrections to the deuteronium
energy levels at order $\alpha^4 m_d$ coming from the Breit Hamiltonian.
Results are given for $n^{2S+1}L_J$ levels with 
$n \in \{ 2, 3,4\}$, $L\ge 1$, and all possible combinations
of $S$ and $J$.  The uncertainties quoted here are due completely to the
uncertainty in $Q_d$, since $r_d$, which also has a relatively large uncertainty,
only enters for $S$ states, and the remaining quantities involved have
negligible uncertainties.  The $n^\pm \! L_L$ states are linear combinations
of $n^1\! L_L$ and $n^5\! L_L$ found by diagonalization of the Breit Hamiltonian
[see the discussion near the end of Sec.~\ref{sec2} and Eq.~\eqref{M211}].}
\renewcommand{\arraystretch}{1.4}
\begin{tabular}{c@{\hspace*{0.5cm}}.@{\hspace*{0.5cm}}c@{\hspace*{0.5cm}}.}
\hline
\hline
$2{}^{2S+1}P_J$ &
\multicolumn{1}{l}{\phantom{xx} $E_{\rm BR}$ (meV) } &
$3{}^{2S+1}P_J$ &
\multicolumn{1}{l}{\phantom{x} $E_{\rm BR}$ (meV) } \\
\hline
$2^{-}\!P_1$ & -551.x52(6)    & $3^{-}\!P_1$ & -169.x06(2)) \\
$2^{+}\!P_1$ & 815.x32(6)    & $3^{+}\!P_1$ & 235.x93(2) \\
$2^{3}P_0$ & -876.x66(6)     & $3^{3}P_0$ & -265.x39(2) \\
$2^{3}P_1$ & 290.x56(3)      & $3^{3}P_1$ & 80.x449(9) \\
$2^{3}P_2$ & -95.x614(6)     & $3^{3}P_2$ & -33.x973(2) \\
$2^{5}P_2$ & -559.x34(4)     & $3^{5}P_2$ & -171.x373(13) \\
$2^{5}P_3$ & 148.x457(12)  & $3^{5}P_3$ & 38.x344(4)  \\
\hline
\hline
$3{}^{2S+1}D_J$ &
\multicolumn{1}{l}{\phantom{xx} $E_{\rm BR}$ (meV) } &
$4{}^{2S+1}P_J$ &
\multicolumn{1}{l}{\phantom{x} $E_{\rm BR}$ (meV) } \\
\hline
$3^{-}\!D_2$ & -49.x419(3) & $4^{-}\!P_1$ & -72.x511(7) \\
$3^{+}\!D_2$ & 18.x051(3) & $4^{+}\!P_1$ & 98.x344(8) \\
$3^{3}D_1$ & -36.x784(2)     & $4^{3}P_0$ & -113.x153(8) \\
$3^{3}D_2$ & 11.x986(2)      & $4^{3}P_1$ & 32.x749(4) \\
$3^{3}D_3$ & -10.x8249(5)   & $4^{3}P_2$ & -15.x5226(8) \\
$3^{5}D_0$ & 19.x590(4)      & $4^{5}P_2$ & -73.x488(6) \\
$3^{5}D_1$ & 1.x603(2)        & $4^{5}P_3$ & 14.x9862(15) \\
$3^{5}D_3$ & -31.x248(2)     &  \\
$3^{5}D_4$ & 10.x9760(10)  &  \\
\hline
\hline
$4{}^{2S+1}D_J$ &
\multicolumn{1}{l}{\phantom{xx} $E_{\rm BR}$ (meV) } &
$4{}^{2S+1}F_J$ &
\multicolumn{1}{l}{\phantom{x} $E_{\rm BR}$ (meV) } \\
\hline
$4^{-}\!D_2$ & -22.x0388(12) & $4^{-}\!F_3$ & -8.x9600(4) \\
$4^{+}\!D_2$ & 6.x4249(11)   & $4^{+}\!F_3$ & 1.x1213(4) \\
$4^{3}D_1$ & -16.x7084(8)   & $4^{3}F_2$ & -6.x2652(2) \\
$4^{3}D_2$ & 3.x8662(8)      & $4^{3}F_3$ & 0.x7084(3) \\
$4^{3}D_3$ & -5.x7571(2)     & $4^{3}F_4$ & -2.x58892(9) \\
$4^{5}D_0$ & 7.x0744(15)    & $4^{5}F_1$ & 0.x0494(5) \\
$4^{5}D_1$ & -0.x5141(8)     & $4^{5}F_2$ & -2.x96259(11) \\
$4^{5}D_3$ & -14.x3730(9)   & $4^{5}F_4$ & -5.x0753(3) \\
$4^{5}D_4$ & 3.x4402(5)      & $4^{5}F_5$ & 1.x4435(2) \\
\hline
\hline
\end{tabular}
\end{center}
\end{table}

%
%
\section{Vacuum Polarization}
\label{sec4}

\subsection{One--Loop Electronic VP}
\label{sec4A}

In momentum space, the Coulomb potential
takes the form $V(\vec k\,) = -4\pi\alpha/\vec k\, ^2$,
where we set the nuclear charge number $Z=1$.
The one-loop correction to
the Coulomb potential, pictured in Fig.~\ref{fig1}(a), is given
by the replacement
\begin{equation}
\label{repl1L}
- \frac{4 \pi \alpha}{\vec k\,^2} \rightarrow
- \frac{4 \pi \alpha}{\vec k\,^2} \, \left [ - \Pi^{(1)}_\rmR(-\vec k\,^2) \right ] \,,
\end{equation}
where $\Pi^{(1)}_\rmR(-\vec k\,^2)$ is the renormalized one-loop
vacuum polarization function in the conventions of Ref.~\cite{LaJe2024},
evaluated for space-like momentum transfer $k^2 = -\vec k\,^2$.
After using a dispersion relation,
one may express the (electronic) vacuum polarization 
potential as
\begin{equation}
\label{Vvp1L}
V^{(1)}_{\rm eVP}(r) = - \frac{\alpha}{\pi} \frac{2}{r} 
\int\limits_{2 m_e}^\infty \dd q
\; \frac{\ee^{- q r}}{q}
\; \mathrm{Im} \! \left[ \Pi^{(1)}_\rmR (q^2 + \ii \epsilon) \right] \, .
\end{equation}
With the substitution~\cite{Sc1970vol3}
\begin{equation}
\label{lambda}
v = \sqrt{1-\frac{4 m_e^2}{q^2}} \, , \quad q = \frac{2 m_e}{\sqrt{1-v^2}} \,,
\end{equation}
one may thus express the one-loop
Uehling potential as follows (see Ref.~\cite{Ue1935}),
\begin{equation}
V^{(1)}_{\rm eVP}(r) = \frac{\alpha}{\pi} \int_0^1 \dd v \, 
f_1(v) \, \ee^{-\lambda_e(v) r} \left ( - \frac{\alpha}{r} \right ) \, ,
\end{equation}
where $\lambda_e(v) \equiv 2m_e/\sqrt{1-v^2}$
and we define the one-loop spectral function as
\begin{equation}
\label{f1}
f_1(v) = \frac{2 v}{1 - v^2}  \,
\frac{{\rm Im} \, \Pi^{(1)}_\rmR( v )}{\alpha} 
= \frac{ v^2 (1-v^2/3) }{1-v^2} \,.
\end{equation}
The energy shift due to this potential is
\begin{equation}
E^{(1)}_{\rm eVP} = \big \langle V^{(1)}_{\rm eVP} \big \rangle = 
\int \dd r \, r^2 \, \left [ R_{n L}(r) \right ]^2 \, V^{(1)}_{\rm eVP}(r) \, .
\end{equation}
There is no dependence on the spin.  Using the non-relativistic wave functions
given in Eq.~\eqref{psi},
one finds that
\begin{equation}
\label{E1VP}
E^{(1)}_{\rm eVP} = \left ( \frac{\alpha}{\pi} \right ) E_0 \, C(n,L) \,,
\end{equation}
where $E_0$ is the deuteronium Rydberg (as given in Table~\ref{table1})
and the dimensionless coefficient $C(n,L)$ is
\begin{multline}
\label{CnL}
C(n,L) = \left ( - \frac{2(n-L-1)!}{n^2 (n+L)!} \right ) 
\int_0^1 \dd v \, f_1(v) 
\\
\! \! \times \! \int_0^\infty \dd \rho \, \rho^{2L+1} 
\ee^{-\rho} \left [ L_{n-L-1}^{2L+1}(\rho) \right ]^2 
\ee^{- n \beta_\Dt \, \rho/\sqrt{1-v^2}} \, ,
\end{multline}
and $\beta_\Dt = m_e/(\alpha m_r)$ has been given 
in Table~\ref{table1}. For given $n$ and $L$,
the $\rho$ integral can easily be performed 
analytically, and the $v$ integral can be 
performed numerically. (We should note that,
for individual states, fully analytic results
exist~\cite{Pu1957,LaJe2024}, but the 
numerical integration proceeds without problems~\cite{Wo1999}
with the built-in numerical integration
routines of modern computer algebra systems.) 
One notes that the potentially singular terms in $1/(1-v^2)$ cancel 
naturally after the integration over $\rho$.
The numerical results are shown in Table~\ref{table4}.
As examples, the $1S$ energy correction is mainly determined by electron vacuum polarization:
\begin{equation}
E^{(1)}_{\rm eVP}(1S) = -125.203\,42 \, {\rm eV} \, ,
\end{equation}
and the $n=3$ Lamb shift due to electron vacuum polarization is
\begin{equation}
E^{(1)}_{\rm eVP} (3P\!-\!3S) = 1.89842 \,{\rm eV} \rightarrow 653.09\,{\rm nm} \, .
\end{equation}
The one-loop eVP shift is by far the dominant
radiative correction in heavy bound 
systems~\cite{Pu1957,OwRe1972,Pa1996mu,JeSoIvKa1997,KaIvJeSo1998,Bo2012,
KaIvKo2012,ReDi2018a,ReDi2018b}.

%
%
\subsection{One--Loop Muonic VP}
\label{sec4B}

In muonic bound systems, one can 
approximate the muonic vacuum polarization potential
by a Dirac delta potential~\cite{Pa1996mu,Je2011aop1},
because the mass scale of the bound particle 
(the muon) is equal to the mass of the 
virtual muon in the vacuum polarization loop.
It means that, for muonic bound systems
and muonic vacuum polarization,
the effective $\beta$ parameter in 
Eq.~\eqref{CnL} is large, approximately equal to $1/\alpha$.
Hence, muonic vacuum polarization effects
are drastically suppressed for bound systems
where the orbiting particle is a muon,
and are in fact of order $\alpha^5 m_r$ for 
$S$ states and of higher order than $\alpha^5 m_r$ 
for non--$S$ states.
It is a characteristic of the 
deuteronium bound system that 
muonic vacuum polarization even makes a
substantial contribution to the 
bound-state energy of non--$S$ states,
because the $\beta$ parameter for 
muonic vacuum polarization in deuteronium,
\begin{equation}
\beta^{\mu}_\Dt = \frac{m_\mu}{\alpha m_r} \approx 15.4
\ll 137.0 \approx 1/\alpha \,,
\end{equation}
is much smaller than the inverse of the fine-structure 
constant. For muonic vacuum polarization,
one replaces, in Eq.~\eqref{CnL},
\begin{equation}
\beta_\Dt \to \beta^\mu_\Dt \,.
\end{equation}
The muonic vacuum polarization potential is 
\begin{equation}
V^{(1)}_{\mu{\rm VP}}(r) = \frac{\alpha}{\pi} \int_0^1 \dd v \,
f_1(v) \, \ee^{-\lambda_\mu(v) r} \left ( - \frac{\alpha}{r} \right ) \,,
\end{equation}
where
\begin{equation}
\label{lambda_mu}
\lambda_\mu(v) = \frac{2 m_\mu}{\sqrt{1-v^2}} \,,
\end{equation}
and the energy shift is
\begin{equation}
\label{E1VPmu}
E^{(1)}_{\mu{\rm VP}} =
\big \langle V^{(1)}_{\mu{\rm VP}}(r) \big \rangle \,.
\end{equation}
Results for the one-loop muonic energy $E^{(1)}_{\rm \mu VP}$
are summarized in Table~\ref{table4}.
While results are strongly suppressed for
states with orbital angular momentum $L \geq 2$,
substantial energy shifts are 
encountered for $S$ and $P$ states.

\begin{table*}
\begin{center}
\caption{\label{table4} Contributions to the Lamb shift in deuteronium in units
of meV.  The second column shows the energy shift due to electron vacuum
polarization (the Uehling shift). The third column gives the contribution of
muon vacuum polarization, the fourth of loop-by-loop reducible electron vacuum
polarization, the fifth of two-loop irreducible electron vacuum polarization,
and the sixth of the second-order perturbation theory with two electron vacuum
polarization potentials.  The ``total VP'' column contains the sum of the
previous four, in meV.  The final three columns give the finite size
correction, the correction due to scalar polarizability, and the strong
interaction contribution.}
\begin{ruledtabular}
\begin{tabular}{c,,,,,,,,@{\hspace{0.5cm}};}
Level 
   & \multicolumn{1}{c}{$E^{(1)}_{\rm eVP}$}
   & \multicolumn{1}{c}{$E^{(1)}_{\mu {\rm VP}}$}
   & \multicolumn{1}{c}{$E^{(2 \rm R)}_{\rm VP}$}
   & \multicolumn{1}{c}{$E^{(2 \rm I)}_{\rm VP}$}
   & \multicolumn{1}{c}{$E^{(1+1)}_{\rm VP}$}
   & \multicolumn{1}{c}{total VP}
   & \multicolumn{1}{c}{$E_{\rm FS}$}
   & \multicolumn{1}{c}{$E_{\rm PS}$}
   & \multicolumn{1}{c}{$E_{\rm S}$} \\
$n L$ 
  & \multicolumn{1}{c}{${\rm (meV)}$} 
  & \multicolumn{1}{c}{${\rm (meV)}$} 
  & \multicolumn{1}{c}{${\rm (meV)}$} 
  & \multicolumn{1}{c}{${\rm (meV)}$} 
  & \multicolumn{1}{c}{${\rm (meV)}$} 
  & \multicolumn{1}{c}{${\rm (meV)}$} 
  & \multicolumn{1}{c}{${\rm (meV)}$} 
  & \multicolumn{1}{c}{${\rm (meV)}$} 
  & \multicolumn{1}{c}{${\rm (meV)}$} \\
\hline\noalign{\smallskip}
$2P$ & -9692.x68328 &  -0.x00331 & -10.x46921 & -60.x10373 & 
  -16.x57279 & -87.x14904 & 5.x83 & -54.x94(12) & 130x(130) \\[3pt]
$3P$ & -2517.x83207 &  -0.x00116 & -2.x86332 & -15.x78409 & 
  -2.x95501 & -21.x60358 & 2.x05 & -18.x09(4) & 46x(46) \\[3pt]
$3D$ & -1274.x84951 &  -0.x00000 & +0.x08013 & -10.x58606 & 
  -1.x18435 & -11.x69028 & 0.x00003 & -0.x7235(15) & 0x.00051(51)  \\[3pt]
$4P$ & -1004.x06887 & -0.x00052 & -1.x22505 & -6.x16306 & 
  -0.x92152 & -8.x31015 & 0.x911 & -7.x898(16) & 20x(20) \\[3pt]
$4D$ & -488.x20600 & -0.x00000 & -0.x05248 & -3.x88547 & 
  -0.x35325 & -4.x29120 & 0.x00002 & -0.x3434(7) & 0x.00031(31) \\[3pt]
$4F$ & -189.x07576 & -0.x00000 & +0.x15319& -1.x97191 & 
  -0.x08324 & -1.x90196 & 0.x00000 & -0.x04906(10) & 0x.00000 \\[3pt]
\end{tabular}
\end{ruledtabular}
\end{center}
\end{table*}

%
%
\subsection{Reducible Two--Loop Correction}
\label{sec4C}

In momentum space, the reducible two-loop correction [see Fig.~\ref{fig1}(b)], 
the loop-after-loop correction, 
can be described by the replacement
of the Coulomb potential according to
[cf.~Eq.~\eqref{repl1L}]
\begin{equation}
\label{repl2L}
- \frac{4 \pi \alpha}{\vec k\,^2} \rightarrow 
- \frac{4 \pi \alpha}{\vec k\,^2} \big [ \Pi^{(1)}_\rmR(-\vec k\,^2) \big ]^2  \,.
\end{equation}
Hence, we can use the formalism used for the one-loop
correction, but with the 
one-loop polarization function $\Pi^{(1)}_\rmR$ squared,
\begin{equation}
V^{(2 \rm R)}(r) = - \frac{\alpha}{\pi} \frac{2}{r} \int\limits_{2 m_e}^\infty \dd q
\; \frac{\ee^{- q r}}{q}
\; \mathrm{Im} \! \left[ 
- \left( \Pi^{(1)}_\rmR (q^2 + \ii \epsilon) \right)^2 \right] \,.
\end{equation}
One may use the identity
\begin{equation}
{\rm Im}\!\left [ \left( \Pi^{(1)}_{\rm R}(q^2 + \ii \epsilon ) \right)^2 \right ] \! =
2 \, {\rm Re} \big [ \Pi^{(1)}_{\rm R}(q^2) \big ] \;
{\rm Im} \big [ \Pi^{(1)}_{\rm R}(q^2 + \ii \epsilon ) \big ]
\end{equation}
above threshold $q^2 > 4 m_e^2$, in order to 
simplify the expression.
We recall from Ref.~\cite{LaJe2024} that
\begin{multline}
\label{Pi1R}
\Pi^{(1)}_\rmR(q^2 + \ii \epsilon) = \frac{\alpha}{\pi} \biggl[
\frac{1}{9} \left(8-3 \tilde v^2\right) \\
-\frac{1}{6} \tilde v \left(\tilde v^2-3\right) \ln
\left(\frac{\tilde v-1}{\tilde v+1}\right) \biggr] \,.
\end{multline}
where
\begin{equation}
\tilde v = \sqrt{ 1 - \frac{4 m_e^2}{q^2 + \ii \epsilon} } \,.
\end{equation}
On the cut ($q^2 > 4 m_e^2$), $\tilde v$ becomes $v+ \ii \epsilon$, where
$v=\sqrt{1-4 m_e^2/q^2}$ satisfies $0<v<1$.
The logarithm in Eq.~\eqref{Pi1R} 
transforms as follows
\begin{equation}
\ln { \left ( \frac{\tilde v-1}{\tilde v+1} \right ) } = 
\ln { \left ( \frac{1-v}{1+v} \right ) } + \ii \pi \, .
\end{equation}
So, on the cut, one has
\begin{equation}
\Im \! \big [ \Pi^{(1)}_\rmR(q^2 + \ii \epsilon) \big ]
= \frac{\alpha v}{6} \left( 3 - v^2 \right) \,.
\end{equation}
The real part, for $q^2 > 4 m_e^2$, is
\begin{equation}
\Re \! \big [ \Pi^{(1)}_\rmR(q^2 + \ii \epsilon) \big ] = \frac{\alpha}{\pi} 
\left[ \frac{8-3 v^2}{9} -
\frac{v^3-3 v}{6} 
\ln\left(\frac{1-v}{1+v}\right) \right] .
\end{equation}
The one-loop spectral function from Eq.~\eqref{f1} is
\begin{equation}
\label{f1alternative}
f_1(v) = \frac{2 v}{1 - v^2}  \,
\frac{{\rm Im} \, \Pi^{(1)}_\rmR( v )}{\alpha} \,.
\end{equation}
For the two-loop reducible diagram,
we can define a corresponding spectral function
\begin{eqnarray}
f_{2 \rm R}(v) &=& \frac{2 v}{1 - v^2}  \, \frac{\pi}{\alpha^2} 
{\rm Im}\!\left [ - \left( \Pi^{(1)}_{\rm R}(q^2 + \ii \epsilon ) 
\right)^2 \right ] 
\\
&\phantom{=} & \hspace{-1.0cm} = -2 f_1(v) 
\left[ \frac{8-3 v^2}{9} -
\frac{v \left(v^2-3\right)}{6} 
\ln\left(\frac{1-v}{1+v}\right) \right] \,. \nonumber \\
\end{eqnarray}
The two-loop reducible potential thus is
\begin{equation}
V^{(2 \rm R)}_{\rm VP}(r) = 
\left( \frac{\alpha}{\pi} \right)^2 \int_0^1 \dd v \,
f_{2 \rm R}(v) \, \ee^{-\lambda_e(v) r} \left ( - \frac{\alpha}{r} \right ) \,,
\end{equation}
and the remaining evaluation of
\begin{equation}
\label{ErVP}
E^{ (2 \rm R)}_{\rm VP} = \left  \langle V^{(2 \rm R)}_{\rm VP}(r) \right \rangle
\end{equation}
proceeds as outlined in Sec.~\ref{sec4A}.
Results for $E_{\rm VP}^{(2 \rm R)}$ are given in Table~\ref{table4}.

%
%
\subsection{Irreducible Two--Loop Correction}
\label{sec4D}

The two-loop irreducible vacuum polarization function
[see Figs.~\ref{fig1}(c) and (d)] was initially calculated
by Kallen and Sabry \cite{KaSa1955} and by
Schwinger \cite{Sc1970vol2}.  A convenient form was
recently obtained in Ref.~\cite{LaJe2024}
for the imaginary part of the two-loop
irreducible vacuum polarization function
above threshold ($q^2 > 4 m_e^2$):
\begin{eqnarray}
\label{ImPi2R}
\frac{\pi}{\alpha^2} {\rm Im} \big [ \Pi^{(2\rm I)}_{\rm R}(q^2 + \ii\epsilon) \big ]
&=& \frac{v(5  -3 v^2)}{8} \nonumber \\
&\phantom{x}& \hspace{-3.2cm} + \frac{1}{48} \left(7 v^4-22 v^2-33\right) \ln\left( -x \right) \nonumber \\
&\phantom{x}& \hspace{-3.2cm} + \frac{v(3 - v^2)}{6} \biggl[ 2 \Phi_1 (x)
+ 3 \ln\left( -x \right) \biggr] \nonumber \\
&\phantom{x}& \hspace{-3.2cm} + \frac{ (v^2-3)(v^2+1)}{6} 
\biggl[ \Phi_1 (x) \ln\left( -x \right) -2 \Phi_2 (x) \biggr] \, ,
\end{eqnarray}
where $x \equiv (v-1)/(v+1)$ and $\Phi_n(x)=\Li_n(x) + 2 \Li_n(-x)$ with $\Li_1(x) \equiv - \ln(1-x)$.
One defines, in analogy with Eq.~\eqref{f1alternative},
\begin{equation}
\label{f2}
f_{2\rm I}(v) = \frac{2 v}{1 - v^2}  \frac{\pi}{\alpha^2}
{\rm Im} \left [ \Pi^{(2\rm I)}_\rmR( v ) \right ] \,.
\end{equation}
The irreducible two-loop potential is thus
\begin{equation}
V^{(2 \rm I)}_{\rm VP}(r) = 
\left ( \frac{\alpha}{\pi} \right )^2 \int_0^1 \dd v \, 
f_{\rm 2I}(v) \, \ee^{-\lambda_e(v) r} \left ( - \frac{\alpha}{r} \right ) \, .
\end{equation}
The remaining evaluation of
\begin{equation}
\label{E2VP}
E^{(2 \rm I)}_{\rm VP} = \langle V^{(2 \rm I)}_{\rm VP}(r) \rangle
\end{equation}
proceeds in full analogy
to the one-loop and two-loop reducible energy shifts.
Results are shown in Table~\ref{table4}.

%
%
\subsection{Second--Order Vacuum--Polarization}
\label{sec4E}

The vacuum-polarization correction also 
makes a contribution in second-order
perturbation theory [see Fig.~\ref{fig1}(e)],
\begin{equation}
E^{(1+1)}_{\rm VP} = 
\langle V^{(1)}_{\rm VP} \, \hat G_{nL} \, V^{(1)}_{\rm VP} \rangle  \,,
\end{equation}
where $\hat G_{nL}$ is the reduced Green function for the 
reference state.  (The reduced Green function has the contribution of the
reference state subtracted out, after which the limit as the energy becomes that of the
reference state is taken.)
Convenient coordinate-space forms for these reduced Green functions have been
given in Ref.~\cite{JoHi1979} in terms of the radial reduced
Green function $\hat g_{n L}$, where
\begin{equation}
\hat G_{nL}(\vec r_1,\vec r_2) = 
\hat g_{n L}(r_1,r_2) \, 
\sum_{m} Y_{L m}(\theta_1,\phi_1) Y^*_{L m}(\theta_2,\phi_2) \, .
\end{equation}
where $\hat g_{n L}$ is radial component, and $L$ 
is the orbital angular momentum of the reference state.
We should remark 
that a couple of corrections are necessary in Ref.~\cite{JoHi1979}: 
$\ln (x)$ in the first line of (2.14) should be $\ln ({x_>})$, 
and $\ee^x$ and $x^k$ in the
last line of Eq.~(2.18) of Ref.~\cite{JoHi1979}
should be $e^{x_<}$ and $({x_<})^k$~\cite{JoPriv2023}.  We
also must multiply $\hat g_{n L}$ by $\alpha m_r^2$ to convert 
from the atomic units used in Ref.~\cite{JoHi1979}
to our natural units. The $\hat g_{n L}$ radial reduced Green functions 
 for $n \in \{1,2,3\}$ were 
also given explicitly in Ref.~\cite{LaFl1977},
and a general expression in terms of Whittaker functions 
for the unreduced radial Green function $g_{nL}$ is given in Sec.~4.3.2
of \cite{JeAd2022book}.  Upon making use of the orthonormality of spherical
harmonics, the energy shift becomes
\begin{multline} 
E^{(1+1)}_{\rm VP} = 
\int \dd r_1 \, \dd r_2 \, r_1^2 \, r_2^2 \, 
R_{n L}(r_1) V^{(1)}_{\rm eVP}(r_1) \hat g_{n L}(r_1,r_2) 
\\
\times
V^{(1)}_{\rm eVP}(r_2) R_{n L}(r_2) \, .
\end{multline} 
It can be written in terms of a dimensionless integral as
\begin{equation} 
E^{(1+1)}_{\rm VP} = \left ( \frac{\alpha}{\pi} \right )^2 E_0 \,
D(n,L) \,, 
\end{equation}
where
\begin{multline} 
D(n,L) = \frac{4}{n} \left ( \frac{p!}{q!} \right )^2 
\int_0^1 \dd u f_1(u) \int_0^1 \dd v f_1(v) 
\\
\times \int_{\rho_2 < \rho_1} \dd \rho_1 \, \dd \rho_2 \, 
(\rho_1 \rho_2)^s \,
\ee^{- [ \rho_1 \omega(u) + \rho_2 \omega(v) ] } \,
\\
\times 
L_p^s(\rho_1) \, h_{n L}(\rho_1,\rho_2) \, L_p^s(\rho_2)
\end{multline} 
where $p \equiv n-L-1$, $q \equiv n+L$, $s \equiv 2L+1$,
\begin{equation}
\omega(v) \equiv 1+n \beta_{\rm Dt}/\sqrt{1-v^2} \, ,
\end{equation}
and
\begin{equation}
\hat g_{n L}(r_1,r_2) = (\alpha m_r^2) \frac{4 p!}{n q!}  (\rho_1 \rho_2)^{L} \, e^{-\frac{\rho_1+\rho_2}{2}} \, h_{n L}(\rho_1,\rho_2) 
\end{equation}
with the usual relations [see Eq.~\eqref{radial_wf}] between the radial variables $r$ and their dimensionless partners $\rho$.
The reduced radial Green functions $h_{n L}(\rho_1,\rho_2)$ 
were obtained from Eq.~(2.18) of \cite{JoHi1979}, being specifically the part of that equation inside the curly bracket
(with the corrections mentioned earlier).
For example,
\begin{multline}
h_{1 0}(\rho_1,\rho_2) = 
\bigg \{ -\frac{7}{2} + 2 \gamma_E-\frac{1}{\rho_1} 
- \frac{1}{\rho_2} \\
+ \frac{\rho_1}{2} + \frac{\rho_2}{2} +\frac{\ee^{\rho_2}}{\rho_2} -
{\rm Ei}(\rho_2)+\ln(\rho_1 \rho_2) \bigg \} \, ,
\end{multline}
which holds when $\rho_2 < \rho_1$.  
The radial integral over $\rho_2$ can be performed
analytically and the integration over $\rho_1$, $v$ and $w$ 
proceeds numerically~\cite{Wo1999}.  
Results for $E^{(1+1)}_{\rm VP}$ are given in Table~\ref{table4}.

%
%
\section{Internal--Structure Effects}
\label{sec5}

%
%
\subsection{Hadronic VP: Calculation}
\label{sec5A}

Even though hadronic VP [see Fig.~\ref{fig1}(f)] constitutes a numerically
small energy correction,
it is an extremely interesting effect to study
as it pertains to the internal hadronic 
structure of the virtual photons
exchanged between the constituent 
particles of deuteronium.
Our approach, as outlined below,
is inspired by Refs.~\cite{JeSoIvKa1997,KaIvJeSo1998}, 
where hadronic VP effects
were studied for true muonium (alternatively 
referred to as dimuonium, the $\mu^+ \mu^-$ bound system).
The strategy will be as follows.
{\em (i)}
The dominant contribution to hadronic 
vacuum polarization comes from the 
pion form factor, 
which has been given in a particularly useful parameterization in Ref.~\cite{GoSa1968}.
{\em (ii)} The $\omega$ and $\phi$ meson contributions 
are taken into account in the pole approximation.
{\em (iii)}
Finally, the high-energy background above 1\,GeV
is estimated by taking into account the 
$R$ ratio of the relevant cross sections of virtual 
photons into hadronic versus light fermionic particles,
in the high-energy region.

First, we deal with the pionic contribution.
The hadronic VP insertion is described
by the following replacement of the momentum space Coulomb potential,
[see Eq.~(2.1) of Ref.~\cite{SaTeYe1984}
and Eq.~(1) of Ref.~\cite{JeSoIvKa1997}], 
\begin{align}
-\frac{4 \pi \alpha}{\vec k^2} \to & \;
-\frac{\alpha}{\pi}\int_{s_0}^\infty
\dd s\,\,\rho_{\rm had}(s)\,\frac{4 \pi \alpha}{\vec k^2 + s} \,,
\end{align}
where $\rho_{\rm had}(s)$ is the hadronic contribution
to the vacuum polarization spectral function.
We assume that the spectral function
$\rho_{\rm had}(s)$ has a sufficiently 
complicated form so that 
analytic evaluations are impossible.
After a transformation to coordinate space,
the hadronic vacuum polarization potential 
takes the form
\begin{equation}
V_{\rm had}(r) = \frac{\alpha}{\pi} \, \left(-\frac{\alpha}{r}\right) \, 
\int_{s_0}^\infty \dd s\,\,\rho_{\rm had}(s)\, \ee^{- \sqrt{s} \, r} \,.
\end{equation}
For the pion insertion, one obtains
\begin{equation}
\label{Vpi}
V_\pi(r) = \frac{\alpha}{\pi} \, \left(-\frac{\alpha}{r}\right) \,
\int_{s_0}^\infty \dd s\,\,\rho_\pi(s)\, \ee^{- \sqrt{s} \, r} \,,
\end{equation}
with $s_0 = (2 m_\pi)^2$.
We are inspired by the approach 
outlined in Ref.~\cite{SaTeYe1984}. The spectral function is of the form 
\begin{equation}
\rho_\pi(s) = \frac{(s - 4\,m_{\pi}^2)^{3/2}}{12\,s^{5/2}}\,
\left | F_{\pi}(s) \right |^2 \,,
\end{equation}
where the pionic form factor is used in the form given by Gounaris and
Sakurai~\cite{GoSa1968}:
\begin{equation}  \label{SGff}
F_{\pi}(s) 
= \frac{N}{D_1 + D_2 - \ii\,D_3}
\approx \frac{N}{D_1 - \ii\,D_3}\, .
\end{equation}
According to Eq.~(11) of Ref.~\cite{GoSa1968},
the pion form factor can 
be approximated in terms
of the quantities $N$, $D_1$, $D_2$ and $D_3$,
as follows,
\begin{subequations}
\begin{align}
N =& \; m_\rho^2 + d \, m_\rho \, \Gamma_\rho
= m_\rho^2 \, \left( 1 + d \, \frac{\Gamma_\rho}{m_\rho} \right) \,,
\\
d =& \; \frac{3 m_\pi^2}{\pi k_\rho^2} 
\ln \left ( \frac{m_\rho + 2 \, k_\rho}{2\,m_\pi} \right ) + 
\frac{m_\rho}{2\,\pi\,k_\rho} - 
\frac{m_\pi^2\, m_\rho}{\pi\,k_\rho^3} \approx 0.48,
\\
D_1 =& \; m_\rho^2 - s \,, \\
D_2 =& \; \Gamma_\rho \frac{m_\rho^2}{k_\rho^3} \,
\bigg[k(s)^2\,\left(h(s) - h_\rho\right) + k_\rho^2 \,\, h'(m_\rho^2) \,\,
(m_\rho^2 - s) \bigg] \,,
\\
D_3 =& \; 
m_\rho\,\Gamma_\rho \, 
\left( \frac{k(s)}{k_\rho} \right)^3 \, \frac{m_\rho}{\sqrt{s}} \,,
\end{align}
\end{subequations}
where $h'$ denotes the derivative of $h$, 
and the functions $k$ and $h$ are defined as
\begin{equation}
k(s)=\frac{1}{2}\sqrt{s - 4m_\pi^2}, \quad h(s) = \frac{2}{\pi} \,
\frac{k(s)}{\sqrt{s}} \, 
\ln \left(\frac{\sqrt{s} + 2\, k(s)}{2\,m_\pi}\right)\,.
\end{equation}
Furthermore, we use the short-hand notations
$k_\rho = k(m_\rho^2)$ and
$h_\rho = h(m_\rho^2)$.
The interpretation is as follows:
$N$ is a normalization factor.
$D_1$ is the dominant term in the 
denominator (in fact, for $D_2 = 0$,
and constant $D_3$, one would obtain 
a Lorentzian resonance profile for the 
$\pi$ resonance). The term $D_2$ corrects
the Lorentzian profile, while $D_3$ parameterizes
the width of the $\pi$ resonance.
We use the following values from the
latest data compilation of the
Particle Data Group (see Ref.~\cite{PDG2024}),
\begin{equation}
\Gamma_\rho = 147.4(8) \, {\rm MeV},
\qquad
m_\rho = 775.26(23) \, {\rm MeV} \,,
\end{equation}
and the charged pion mass 
$m_\pi = 139.57039(18) \, {\rm MeV}$.

Numerically, the contribution of the
$D_2$ is smaller than that of $D_1$.
Hence, we can use the following approximation,
\begin{equation}
\label{form_factor_approximation}
| F_\pi(s) |^2 = \frac{N^2}{(D_1 + D_2)^2 + (D_3)^2}
\approx \frac{N^2}{(D_1)^2 + (D_3)^2} \,.
\end{equation}
The difference of the results obtained
with the approximation $D_2 = 0$, and without,
is taken as the basis for an estimate
of the theoretical uncertainty
of the pionic contribution to hadronic vacuum polarization.
The same approach was applied in Ref.~\cite{SaTeYe1984}.
However, one should note the following
necessary corrections to Eq.~(2.12) of Ref.~\cite{SaTeYe1984},
\begin{subequations}
\begin{align}
\left( 1 + \frac{\Gamma_\rho^d}{m_\rho} \right)^2 \to & \;
\left( 1 + d \frac{\Gamma_\rho}{m_\rho} \right)^2 \,,
\\
\left( \frac{s - 4 m_\pi^2}{m_\rho^2 - 4 m_\pi^2} \right)^{3/2} \to & \;
\left( \frac{s - 4 m_\pi^2}{m_\rho^2 - 4 m_\pi^2} \right)^3 \,.
\end{align}
\end{subequations}
The relative uncertainties based on this error estimate 
range from about 3\% (for $S$ states) to 11\% (for the $4F$ state) 
for principal quantum numbers $n=1,2,3,4$.

We can now proceed to evaluate
\begin{equation}
\label{E_pi}
E_\pi(nL) = \big \langle V_\pi(r) \big \rangle \, =
\int \dd r \, r^2 \, \left [ R_{n L}(r) \right ]^2 \, V_\pi(r) \, .
\end{equation}
For example, for the $2P$ state of deuteronium, one obtains the results
\begin{equation}
E_{\pi}(2P) = -4.1(2) \times 10^{-7} \, {\rm eV} \,,
\end{equation}
which constitutes the dominant hadronic VP correction
to the energy of this state.

For $\omega$ and $\phi$ mesons,
it is adequate to use the pole approximation,
\begin{equation}
\label{analytic1}
\rho_p(s) = \frac{4\pi^2}{f_p^2} \, \delta(s - m_p^2) \,,
\end{equation}
where the subscript $p$ designates the particle being considered.
After Fourier transformation, one can express
the vacuum polarization potential due to the resonance 
with particle $p$ as a Yukawa potential,
with a specific coupling parameter,
\begin{equation}
V_p(r) = -\frac{\alpha}{\pi} \frac{4\pi^2}{f_p^2} \,
\frac{\alpha}{r} \, \ee^{-m_p r} \,.
\end{equation}

We can use the parameters
\begin{equation}
\frac{f_\omega^2}{4\pi} = \; 18(2) \,, \qquad
\frac{f_\phi^2}{4\pi} = \; 11(2) \,
\end{equation}
(see Refs.~\cite{SaTeYe1984,BaSpYePi1978})
and the mass values $m_\omega = 782.66(13) \, {\rm MeV}$, 
$m_\phi = 1019.461(16) \, {\rm MeV}$ 
of Refs.~\cite{TiMoNeTa2021,MoNeTaTi2025}.
The hadronic vacuum-polarization potential $V_{p}(r)$
due to one of the $\omega$ and $\phi$ mesons 
leads to the energy shift
\begin{equation}
\label{E_omegaphi}
E_{p}(nL) = 
\big \langle V_{p}(r) \big \rangle \,
= \int \dd r \, r^2 \, \left [ R_{n L}(r) \right ]^2 \,V_{p}(r) \, .
\end{equation}

The contribution
of the high-energy background above $1\,{\rm GeV}$ can be
estimated, in the dispersive approach,
on the basis of the spectral function
\begin{equation}
\label{analytic2}
\rho(s) = \frac{R}{3 s}, \qquad
m_1 < \sqrt{s} < m_2 \,,
\end{equation}
where $R$ is the ratio of a virtual
photon going into a hadronic pair,
to a virtual photon going into a
light fermion pair~\cite{JeSoIvKa1997}.
Assuming $R$ to be piecewise constant is a permissible
approximation, because, according
to Ref.~\cite{JeSoIvKa1997},
\begin{align}
\label{Rapprox}
R \approx & \;
\left\{
\begin{array}{cc}
R_2 = 2 \;\; & \quad m_1 \equiv
1\,{\rm GeV} < \sqrt{s} < m_2 \equiv 4\,{\rm GeV} \\[2ex]
R_4 = 4 \;\; & \quad m_2 \equiv
4\,{\rm GeV} < \sqrt{s} < \infty
\end{array}
\right. \,.
\end{align}
According to Fig.~18 of Ref.~\cite{KeNoTe2018},
the approximation~\eqref{Rapprox} still represents
a suitable approximation to the additional,
recently  available experimental data in the relevant
energy domain.
Hence, the vacuum polarization potential
describing the high-energy background 
for $m_1 < \sqrt{s} < m_2$ 
is (in momentum space)
\begin{equation}
V_{m_1,m_2}(\vec k) = -\frac{\alpha}{\pi}\int_{(m_1)^2}^{(m_2)^2}
\dd s\,\,\frac{R}{3s}\,\frac{4 \pi \alpha}{\vec k^2 + s} \,.
\end{equation}
After Fourier transformation, the potential becomes
\begin{multline}
V_{m_1,m_2}(r) = \frac{\alpha}{\pi} \, \left ( - \frac{\alpha}{r} \right )
\frac{2 R }{3} \,
\Bigl [ {\rm Ei}(-m_2 r) - {\rm Ei}(-m_1 r) \Bigr ] \,.
\end{multline}
Here, the exponential integral ${\rm Ei}(z)$ is given as follows,
\begin{equation}
{\rm Ei}(z) \equiv - \int_{-z}^\infty \frac{\dd t}{t} e^{-t} \, .
\end{equation}
The hadronic vacuum polarization potential 
due to the high-energy background
can thus be approximated as 
\begin{eqnarray}
\label{Vb}
V_{b}(r) &=& \frac{\alpha}{\pi} \, \left ( - \frac{2 \alpha}{3 r} \right ) \,
\bigg[  R_2 \Big\{ {\rm Ei}(-m_2 r) - {\rm Ei}(-m_1 r) \Bigr \}
\cr
&\phantom{x}& 
\hspace{1.9cm} + R_4 \Bigl\{  0 - {\rm Ei}(-m_2 r) \Bigr\} \bigg] \,,
\end{eqnarray}
leading to the energy shift
\begin{equation}
\label{E_b}
E_{b}(nL) = 
\big \langle V_{b}(r) \big \rangle \,
= \int \dd r \, r^2 \, \left [ R_{n L}(r) \right ]^2 \,V_{b}(r) \, ,
\end{equation}
which can easily be evaluated numerically.

According to Eqs.~\eqref{E_pi},~\eqref{E_omegaphi}, and~\eqref{E_b},
the result for the hadronic vacuum polarization energy shift
$E_{\rm had}(r)$ is 
\begin{equation}
E_{\rm had}(nL) = E_\pi(nL) + E_{\omega}(nL) + E_{\phi}(nL) + E_b(nL) \,.
\end{equation}
For the states of interest for the current investigation,
one obtains
\begin{subequations}
\label{E_had}
\begin{align}
E_{\rm had}(2P) =& \;  -4.8(5) \times 10^{-4} \, {\rm meV} \,,
\\
E_{\rm had}(3P) =& \;  -1.7(2) \times 10^{-4} \, {\rm meV} \,,
\\
E_{\rm had}(3D) =& \;  -2.9(3) \times 10^{-9} \, {\rm meV} \,,
\\
E_{\rm had}(4P) =& \;  -7.5(8) \times 10^{-5} \, {\rm meV} \,,
\\
E_{\rm had}(4D) =& \;  -1.8(2) \times 10^{-9} \, {\rm meV} \,,
\\
E_{\rm had}(4F) =& \;  -1.8(2) \times 10^{-14} \, {\rm meV} \,.
\end{align}
\end{subequations}
We have included an uncertainty of about 10\%, as suggested by the effect of
the approximation \eqref{form_factor_approximation} in the pion contribution,
the simplicity of our approximations, and the size of the error in the estimate
of the hadronic VP contribution to the muon anomalous moment, as
described below.  These energy corrections are small compared to other
contributions of order $\alpha^3$ or $\alpha^4$.

%
%
\subsection{Hadronic VP: Verification}
\label{sec5B}

In order to verify our approximate approach for the 
treatment of hadronic vacuum-polarization, we have carried
out a calculation of the corresponding 
contribution to the muon anomalous magnetic moment $a_\mu$.
The leading two-loop hadronic
contribution to the muon anomalous
magnetic moment is~\cite{LaPedR1972,Ad1974prd1,Je2011aop1}
\begin{equation}
\label{a_mu_had}
\delta a_{\mu, {\rm had}} =
\left( \frac{\alpha}{\pi} \right)^2 \,
\int_{s_0}^\infty \dd s \,
K_\mu(s) \, \rho_{\rm had}(s) \,,
\end{equation}
where the kernel function is
\begin{equation}
\label{a_mu_kernel}
K_\mu(s) = \int_0^1 \dd x \,
\frac{x^2 \, (1-x)}{x^2 + (1-x) \, s/m_\mu^2} \,.
\end{equation}
As a check of the formulas, we consider the replacements
$m_\mu \rightarrow m_e$ and
\begin{equation}
\rho_{\rm had}(s) \rightarrow \rho_e(s) = \frac{1}{\alpha s} {\rm Im} \left [ \Pi^{(1)}_R(s) \right ]
= \frac{v (1-v^2/3)}{2s} \, 
\end{equation}
in Eqs.~\eqref{a_mu_had} and \eqref{a_mu_kernel}.  With those replacements, one obtains the
known contribution to the electron anomalous moment due to one-loop electron vacuum
polarization (see p.~852 of Ref.~\cite{BaMiRe1972a}),
\begin{equation}
\delta a_{e,{\rm eVP}} =
\left( \frac{\alpha}{\pi} \right)^2
\left( \frac{119}{36} - \frac{\pi^2}{3} \right)  \,.
\end{equation}

We use the formula
$\rho_{\rm had}(r) = \rho_\pi(r) + \rho_\omega(r) + \rho_\phi(r) + \rho_b(r)$
for the spectral density of hadronic vacuum polarization and
Eq.~\eqref{a_mu_had} to obtain the hadronic VP contribution to the muon anomalous moment. We find
\begin{align}
\label{a_mu_result}
a_{\mu, {\rm had}} =& \; 
\underbrace{4.39 \times 10^{-8}}_{\pi^+ \, \pi^-} + 
\underbrace{0.50 \times 10^{-8} }_{\mbox{$\omega$ meson}}
\nonumber\\
& \; + \underbrace{0.51 \times 10^{-8}}_{\mbox{$\phi$ meson}} + 
\underbrace{1.35 \times 10^{-8} }_{\mbox{background}}
\nonumber\\
= & \; 6.75 \times 10^{-8} \,.
\end{align}
The difference between this value and the value of 
$a_{\mu, {\rm had}} = 6.93 \times 10^{-8}$
given in Ref.~\cite{KeNoTe2018}
is roughly 3\,\%. The high degree of agreement between our result and the
result of more careful evaluations supports the 
assumption of the general validity of our 
simplified approach (and the smallness of the corresponding energy correction).

\renewcommand{\arraystretch}{1.4}
\begin{table*}[t!]
\begin{minipage}{0.9\linewidth}
\caption{\label{table5} Finite-size and strong-interaction energy shifts for
deuteronium states.  Shown are the exact energy shifts [see Eq.~\eqref{Eexact}]
between a pointlike
Coulomb system and a system with one particle being a uniformly charged sphere
(doubled because both particles have finite size), the same energy shifts
calculated perturbatively [see Eq.~\eqref{Epert}], 
the Breit energy shifts $E_{\rm D}$ proportional to
$r_d^2$ [see Eq.~\eqref{Ediracdelta}], and the strong interaction energy shifts $E_{\rm S}$
[see Eq.~\eqref{trueman}].  The
uncertainties in the strong interaction shifts are estimated to be $100 \%$ of
their value.  All energies are given in meV.}
\begin{tabular}{lSSSS}
\hline
\hline\noalign{\smallskip}
Level 
  & \multicolumn{1}{c}{$E_{\rm FS}^{\rm exact}$} 
  & \multicolumn{1}{c}{$E_{\rm FS}^{\rm pert}$} 
  & \multicolumn{1}{c}{$ E_{\rm D}$} 
  & \multicolumn{1}{c}{$ E_S$} \\
$n L$ 
  & \multicolumn{1}{c}{${\rm (meV)}$} 
  & \multicolumn{1}{c}{${\rm (meV)}$} 
  & \multicolumn{1}{c}{${\rm (meV)}$} 
  & \multicolumn{1}{c}{${\rm (meV)}$} \\
\hline\noalign{\smallskip}
$1S$ & 3.19e5   & 3.37e5    & 3.63e5 & 3.46e6 \\
\noalign{\smallskip}
$2S$ & 4.00e4   & 4.21e4    & 4.53e4 & 4.33e5 \\
$2P$ & 5.83     & 5.87      & 0      & 130 \\
\noalign{\smallskip}
$3S$ & 1.19e4   & 1.25e4    & 1.34e4 & 1.28e5 \\
$3P$ & 2.05     & 2.06      & 0      & 45.7 \\
$3D$ & 3.00e-5  & 3.01e-5   & 0      & 5.09e-4 \\
\noalign{\smallskip}
$4S$ & 5.00e3   & 5.26e3    & 5.67e3 & 5.41e4 \\
$4P$ & 0.911    & 0.918     & 0      & 20.3 \\
$4D$ & 1.80e-5  & 1.81e-5   & 0      & 3.06e-4 \\
$4F$ & 5.73e-11 & 5.74e-11  & 0      & 5.11e-10 \\
\hline
\hline
\end{tabular}
\end{minipage}
\end{table*}

%
%
\subsection{Finite--Size Effect}
\label{sec5C}

The deuteron has a finite size; its root-mean-square (RMS)
charge radius is $r_d = 2.12778(27) \,
{\rm fm}$. So, the interaction between deuteron and antideuteron is not really
that between two point-like particles.  The fact that the charge is spread out
means that the average potential will not 
be quite as negative as it would be were
the particles pointlike, so the binding energy will be less in magnitude that
it would be for pointlike particles.  That is, the energy correction due to the
finite size of the particles will be positive.  We can make an estimate of the
size of this correction by calculating the interaction energy for two uniformly
charged spheres of radius $r_d$. This estimate is fully sufficient
for our purposes, because the nuclear-structure effects
are numerically dominated by the strong-interaction correction
discussed in Sec.~\ref{sec5E}.

Consider first the case of a pointlike particle of mass $m_1$ and charge
$q_1=-e$ in the field of a uniformly charged spherical shell
of radius $r_d$, mass $m_2$, and charge $q_2=e$. 
The Schr\"odinger equation describing this situation is
\begin{equation}
- \frac{1}{2 m_r} \nabla^2 \psi(\vec r\,) + V(r) \psi(\vec r\,) = E \psi(\vec r\,) \, ,
\end{equation}
where
\begin{equation}
V(r) = \begin{cases} - \dfrac{\alpha}{r_d} & {\rm if} \; r<r_d \\[4ex]
- \dfrac{\alpha}{r} & {\rm if} \;  r>r_d \end{cases} \, .
\end{equation}
We write $\psi(\vec r\,) = R_{\nu L}(r) \, Y_{L m}(\theta,\varphi)$ 
and $R_{\nu L}(r) = \frac{u_{\nu L}(r)}{r}$. 
In terms of $u_{\nu L}(r)$, the radial equation is
\begin{multline}
- \frac{1}{2 m_r} 
\frac{\dd^2}{\dd r^2} u_{\nu L}(r) + 
\left ( V(r) + \frac{L(L+1)}{2 m_r r^2} \right ) u_{\nu L}(r) \\
= E_\nu \, u_{\nu L}(r) \, .
\end{multline}
We define $\nu$ so that the energy eigenvalue is 
$E_\nu = - \frac{m_r \alpha^2}{2 \nu^2}$, 
and define a scaled radial variable $\rho$ with
\begin{equation}
\rho = \sqrt{-8 \, m_r \, E_\nu} \, r = \frac{2 m_r \alpha}{\nu} r \, .
\end{equation}
In terms of $\rho$, the sphere boundary is at $\rho_d = \frac{\kappa}{\nu}$, 
with $\kappa=\frac{2 r_d}{a_0}$ where $a_0=\frac{1}{m_r \alpha}$ is the Bohr radius. 
Numerically, one has $a_0 = 28.834\,200\,570(10)$ 
(see Table~\ref{table1}) and $\kappa=0.147587$.
For $\rho < \rho_d$, the radial equation becomes
\begin{equation} 
\label{small_rho_Sch}
\left ( \frac{\dd^2}{\dd \rho^2} 
  - \frac{L(L+1)}{\rho^2} + \frac{\nu^2}{\kappa} - \frac{1}{4} \right ) 
  u_{\nu L}(\rho) = 0 \, ,
\end{equation}
and for $\rho > \rho_d$ it is
\begin{equation} \label{large_rho_Sch}
\left ( \frac{\dd^2}{\dd \rho^2}  - \frac{L(L+1)}{\rho^2} + 
  \frac{\nu}{\rho} - \frac{1}{4} \right ) u_{\nu L}(\rho) = 0 \, .
\end{equation}
The solution for Eq.~(\ref{small_rho_Sch}) that is finite as $\rho \rightarrow 0$ is
\begin{equation}
u_{\nu L}(\rho) = \sqrt \rho \, 
  J_{L+1/2}\left ( \rho \sqrt{\frac{\nu^2}{\kappa}-\frac{1}{4} } \right ) \, ,
\end{equation}
where $J_\ell(r)$ is a Bessel function of the first kind,
and the solution for Eq.~(\ref{large_rho_Sch}) that 
is finite as $\rho \rightarrow \infty$ is
\begin{equation}
u_{\nu L}(\rho) = W_{\nu,L+1/2}(\rho) \, ,
\end{equation}
where $W_{\nu,\ell}(r)$ is a Whittaker $W$ function.  Continuity of the wave
function and its first derivative at $\rho = \rho_d$ requires the continuity of
the derivative of the logarithm $\ln[u_{\nu L}(\rho)]$.  Upon solving this continuity
equation for a given $L$ and for $\nu \approx n$, one finds the appropriate
value of $\nu$ to use in the energy $E_\nu = - \frac{m_r \alpha^2}{2 \nu^2}$.  The
finite-size energy correction is thus
\begin{equation}
\label{Eexact}
E_{\rm FS}^{{\rm exact}} = E_\nu - E_n = 
  - \frac{m_r \alpha^2}{2} \left ( \frac{1}{\nu^2} - \frac{1}{n^2} \right ) \, .
\end{equation}
For deuteronium, this expression must be doubled 
since both deuteron and antideuteron have extension.  

One could also approach the finite size correction perturbatively, 
with a perturbing potential
\begin{equation}
\label{Epert}
\delta V(r) = \left ( -\frac{\alpha}{r_d} + \frac{\alpha}{r} \right ) \,
\Theta(r < r_d) \,,
\end{equation}
where $\Theta$ is the Heaviside step function.
Results for the perturbative shifts $E_{\rm FS}^{\rm pert}$ are given in Table~\ref{table5}.

We have already obtained a ``finite-size'' effect in our calculation of the
Breit correction, namely the $r_d$ term in $E_{\rm BR}$.  This term is
\begin{equation}
\label{Ediracdelta}
E_{\rm D} = \frac{\alpha^4 \, m_d \, \tilde r_d^2}{6 n^3} \delta_{L,0}
\end{equation}
where $\tilde r_d = m_d r_d/(\hbar c) \approx 20.2248(26)$ [see
Eq.~\eqref{tilderd}].  The Breit finite-size correction $ E_{\rm D}$ vanishes
for $L>0$.  The calculation above shows how the extended deuteron charge
distribution affects states with $L>0$.  Due to the small size of $r_d/a_0$,
which equals $0.0738$, the finite size correction decreases rapidly as $L$
increases.

Numerical results for deuteronium states with $n$ = 1, 2, 3, 4 are shown in
Table~\ref{table5}.  The numerical values obtained from the perturbative
calculation ($ E_{\rm FS}^{\rm pert}$) are consistent with the exact results
from actually solving the differential equation and finding the energy
eigenvalue ($ E_{\rm FS}^{\rm exact}$) for all states.  The Breit contribution
$ E_{\rm D}$ is roughly consistent with those two for $S$ states, but misses the
small corrections for non-$S$ states.

%
%
\subsection{Deuteron Polarizability}
\label{sec5D}

The deuteron can be deformed by an external electric field.  The extent
of this deformation is described by polarizability constants.
For  the (static) scalar and tensor polarizabilities of the deuteron,
we use the calculated results of Friar and
Payne~\cite{FrPa2005deuteron},
\begin{subequations} 
\label{alphaEtaud}
\begin{eqnarray}
\alpha_E &=& 0.6330(13) \, {\rm fm}^3 \, , \label{calc_alphaE} \\
\tau_d &=& 0.0317(3) \, {\rm fm}^3 \,.
\end{eqnarray} 
\end{subequations}
These constants enter the polarizability tensor 
$(\alpha_P)^{i j}$ of the deuteron,
which is expressed as 
\begin{equation}
(\alpha_P)^{i j} = \alpha_E \frac{\delta^{i j}}{3} +
\ii \, \sigma_d \, \epsilon^{i j k}
\frac{\genspin^k}{2} + \tau_d \left ( \genspin^i \genspin^j  \right )^{(2)} \,.
\end{equation}
Here, 
the constants $\alpha_E$, $\sigma_d$ and $\tau_d$ parameterize the
electric scalar ($\ell=0$), vector ($\ell = 1$) and tensor ($\ell = 2$) 
components of the deuteron's polarizability, respectively.
These results are expressed in terms of a polarization
volume, {\em i.e.}, in units of fermi cubed.
Temporarily restoring SI units according to standard 
procedures~\cite{Ja1998}, we indicate the conversion of the 
polarization volume to the (static) SI polarizability,
which, for the scalar polarizability, reads as follows,
\begin{equation}
\label{alphaSI}
\alpha_{E,{\rm SI}} \,
= 4 \pi \epsilon_0 \, \alpha_{E,{\rm vol}} \,,
\end{equation} 
where $\epsilon_0$ is the vacuum permittivity.
For a recent overview of the use of the SI polarizability
(units are those of a dipole moment squared divided by 
an energy) in a related context, see Ref.~\cite{Je2024multipole}.
The numerical values in Eq.~\eqref{alphaEtaud}
thus have to be interpreted in terms of volume 
polarizabilities.

Measurements of the deuteron scalar polarizability were made by two groups and
date a few decades back.  The first is by Rodning {\it et al.} in 1982 (see
Ref.~\cite{RoEtAl1982}).  They give references to six prior theoretical
calculations of the deuteron polarizability with values ranging from $0.21 \,
{\rm fm}^3$ to $0.64 \, {\rm fm}^3$, with a best estimate close to $0.60 \,
{\rm fm}^3$.  They measured a value
(from now on, we assume that all polarizabilities 
are indicated in the ``volume'' conventions)
\begin{equation}
\alpha_{E} = 0.70(5) \, {\rm fm}^3 \, ,
\end{equation}
based on elastic scattering of deuterons on lead.  A second experimental value
was reported by Friar {\it et al.} in 1983 (see Ref.~\cite{FrEtAl1983}), 
based on deuteron
photoabsorption data, a dispersion relation and other theoretical input, and a
number of corrections.  Their result was
\begin{equation}
\alpha_{E} = 0.64(5) \, {\rm fm}^3 \, .
\end{equation}
A weighted average of these two values is
\begin{equation}
\alpha_{E} = 0.645(32) \, {\rm fm}^3 \, ,
\end{equation}
which is consistent with the calculated value 
of Eq.~\eqref{calc_alphaE}. To the best of our knowledge,
there are currently no experimental values
available for the deuteron's tensor polarizability.

A comprehensive derivation of the polarizability 
correction of an atomic core for non-$S$ states
of a bound system is presented
in Sec.~6.6 of Ref.~\cite{JeAd2022book}.
Because the focus in the current work is on $S$ states,
we adopt this treatment here.
A  generalization of 
Eq.~(6.177) of Ref.~\cite{JeAd2022book} to account for the
tensor structure of the polarizability, and also
accounting for the two polarizable 
constituent particles, allows us to find the 
polarizability contribution
to the interaction Hamiltonian
\begin{eqnarray}
H_P &=& -\frac{\alpha}{2} \frac{\hat x^i}{r^2} 
\left [ 3 \alpha_P^{i j}(d) 
+ 3 \alpha_P^{i j}(\bar d) \right ] \frac{\hat x^j}{r^2} 
\nonumber \\
&\phantom{=}& \hspace{-0.8cm} = 
- \frac{\alpha}{2} \! \left \{ 2 \alpha_E \delta^{i j} 
\!+\! 3 \tau_P \big [ (S_+^i S_+^j )^{(2)} 
\!+\! (S_-^i S_-^j )^{(2)} \big ] \right \} 
\frac{\hat x^i \hat x^j}{r^4} \nonumber \\
&\phantom{=}& \hspace{-0.8cm} = - \alpha_E \frac{\alpha}{r^4} 
- \frac32 \tau_P \big [ (S_+^i S_+^j )^{(2)} \!+\! (S_-^i S_-^j )^{(2)} \big ] 
(\hat x^i \hat x^j)^{(2)} \frac{\alpha}{r^4} 
\nonumber \\
&\phantom{=}& \hspace{-0.8cm} \equiv H_{\rm PS} + H_{\rm PT} \, .
\end{eqnarray}
Here, $H_{\rm P}$ is the sum of scalar 
($H_{\rm PS}$) and tensor ($H_{\rm PT}$) components.  
Throughout the above derivation, 
we have converted the static polarizability
$\alpha(\omega \!=\! 0)$ used in
Eq.~(6.177) of Ref.~\cite{JeAd2022book}
to a volume polarizability by dividing by $4 \pi$.

The energy correction due to the scalar polarizability factor is
\begin{eqnarray}
\label{EPS}
E_{\rm P S} &=& - \alpha_E \left \langle \frac{\alpha}{r^4} \right \rangle \nonumber \\
&\phantom{=}& \hspace{-1.4cm} 
= -\frac{\tilde \alpha_E E_0 \alpha^3 
[ 3n^2-L(L+1) ]}{L (L+1)(2L-1)(2L+1) (2L+3) n^5} \, ,
\end{eqnarray}
where $E_0$ is given in Table~\ref{table1},
and $\tilde \alpha_E$ is the dimensionless scalar polarizability factor,
\begin{equation} \label{tilde_alphaE}
\tilde \alpha_E 
= m_d^3 \, \alpha_E = \frac{(m_d \, c^2)^3 \alpha_E}{(\hbar c)^3} = 543.6(1.1) \, .
\end{equation}
We used Eq.~(4.346d) of Ref.~\cite{JeAd2022book} for the expectation value of
$1/r^4$.  Results for the scalar polarizability energy corrections are
tabulated in Table~\ref{table4}.  Due to the large size of $\tilde \alpha_E$,
the order-$\alpha^5$ scalar polarizability contributions are comparable in size
to the $\alpha^4$ contributions listed in Table~\ref{table4} for the states
considered.  

The matrix elements of the tensor polarizability Hamiltonian involve the
$D_{LS'SJ}$ angular factor found in \eqref{def_DLSJ}.  One has
\begin{eqnarray}
(H_{\rm PT})_{n LS'SJ} &\phantom{x}& \nonumber \\
&\phantom{=}& \hspace{-2.4cm} 
= -  \frac{3 \tilde \tau_P E_0 \alpha^3 D_{L S' S J} \, 
  [ 3n^2-L(L+1) ]}{2 L (L+1)(2L-1)(2L+1) (2L+3) n^5} \, ,
\end{eqnarray}
where the dimensionless tensor polarizability factor is
\begin{equation} \label{tilde_tauE}
\tilde \tau_P = m_d^3 \, \tau_P = 
  \frac{(m_d \, c^2)^3 \tau_P}{(\hbar c)^3} = 27.2(0.3) \, .
\end{equation}

The tensor polarizability shifts (using the calculated value of $\tau_P$) are
relatively small compared to the Breit energy corrections due to the extra
factor of $\alpha$.  For example, the shifts for the $4^3 D_J$ levels are
\begin{subequations} 
\begin{alignat}{2}
E_{\rm PT}(4^3 D_1) &= -&&0.00860(8) \, , \\
E_{\rm PT}(4^3 D_2) &=  &&0.00860(8) \, , \\
E_{\rm PT}(4^3 D_3) &= -&&0.00246(3) \, ,
\end{alignat} 
\end{subequations}
which are small compared to the corresponding Breit shifts from
Table~\ref{table3}.  Contributions of this size should be noticeable in the
data given precise enough measurements of the $4F \rightarrow 4D$ transition
energies.

%
%
\subsection{Strong Interaction Correction}
\label{sec5E}

Just as in protonium~\cite{KlBaMaRi2002}, the strong interaction 
modifies the deuteron-antideuteron
interaction at short distances of about 1\,fm (the effective range of the strong interaction) in a way that is easily estimated. 
One can assume that the probability density vanishes for distances
$r \lesssim 1\,$fm due to
strong interaction mediated annihilation 
(see the comprehensive discussion
in Sec.~3.12 of Ref.~\cite{KlBaMaRi2002} for protonium).
Hence, we treat the problem
as if there was a hard cutoff at $a = a_{\rm had} = 1\, {\rm fm}$.
We estimate the effect based on the Deser--Trueman
formula~\cite{DeGoBaTh1954,BedH1955,By1957,Tr1961,KlBaMaRi2002,AdJe2025trueman},
for which a suitable generalization to 
excited states has recently been found in Ref.~\cite{AdJe2025trueman}.
We find~\cite{AdJe2025trueman} that the non-relativistic energy levels
in the cut-off potential are shifted from their pure Coulomb
values by the positive amount
\begin{equation}
\label{trueman}
E_S =
\frac{2 \alpha_{n L} \, \beta_L}{n^3} \left ( \frac{a}{a_0} \right )^{2L+1} \,
\alpha^2 \, m_r \,,
\end{equation}
where $a_0$ is the deuteronium Bohr radius given in Table~\ref{table1}, and
\begin{equation}
\alpha_{n L} = \prod_{s=1}^L \left ( \frac{1}{s^2}-\frac{1}{n^2} \right ) \, ,
\quad \beta_L = \frac{2L+1}{\left[ (2L+1)!! \right]^2} \, 
\end{equation}
with $\alpha_{n0}\equiv 1$.
Corresponding results are entered in Tables~\ref{table4} and \ref{table5}.
We conservatively estimate their uncertainty
as 100\% of the obtained results. 
The $S$-wave scattering length generalizes the concept of the hard-sphere 
radius $a_h$ for general electromagnetically bound systems with 
orbiting hadrons~\cite{DeGoBaTh1954,BedH1955,By1957,Tr1961,KlBaMaRi2002,AdJe2025trueman}.
Our estimate, given in Eq.~\eqref{trueman},
is supported by relatively recent investigations which
indicate that the scattering length, as compared to 
protonium, is a decreasing function 
of the atomic weight (see the conclusions of Ref.~\cite{PrBoLRZe2000}).
The strong-interaction shifts follow the same general
pattern as a function of $L$ as the finite-size corrections of the previous section, 
and are even more drastically suppressed for higher angular 
monenta in view of the squared double factorial $[ (2L+1)!! ]^2$
in the denominator of Eq.~\eqref{trueman}.
Still, for typical states under investigation
here, they are larger than the finite-size corrections by an order of magnitude.

%
%
\section{Dipole Transitions}
\label{sec6}

%
%
\subsection{General Properties of the Spectrum}
\label{sec6A}

Our investigations have revealed
that in deuteronium, a bound system with two spin-1 
particles, the spin-dependent (hyper)fine structures
are composed very differently in comparison
to bound systems where the orbiting particles 
have half-integer spin ({\em e.g.}, positronium).
However, there are also similarities:
Namely, just as in positronium, the spin-dependent
corrections in deuteronium all have the 
same order-of-magnitude ($\alpha^4 m_d$).
We have studied, in Sec.~\ref{sec4},
the leading radiative corrections
(one-loop and two-loop eVP) for the lowest
non-$S$ levels in the bound system.
For states with the same principal 
quantum number $n$, the dominant radiative correction
(one-loop eVP) to the energy 
is more strongly negative (hence, more attractive) 
for states low orbital angular momenta, leading to the 
dominant contribution
to the Lamb shift in the bound system.

\begin{figure}[t!]
\begin{center}
\begin{minipage}{0.99\linewidth}
\begin{center}
\includegraphics[width=0.91\linewidth]{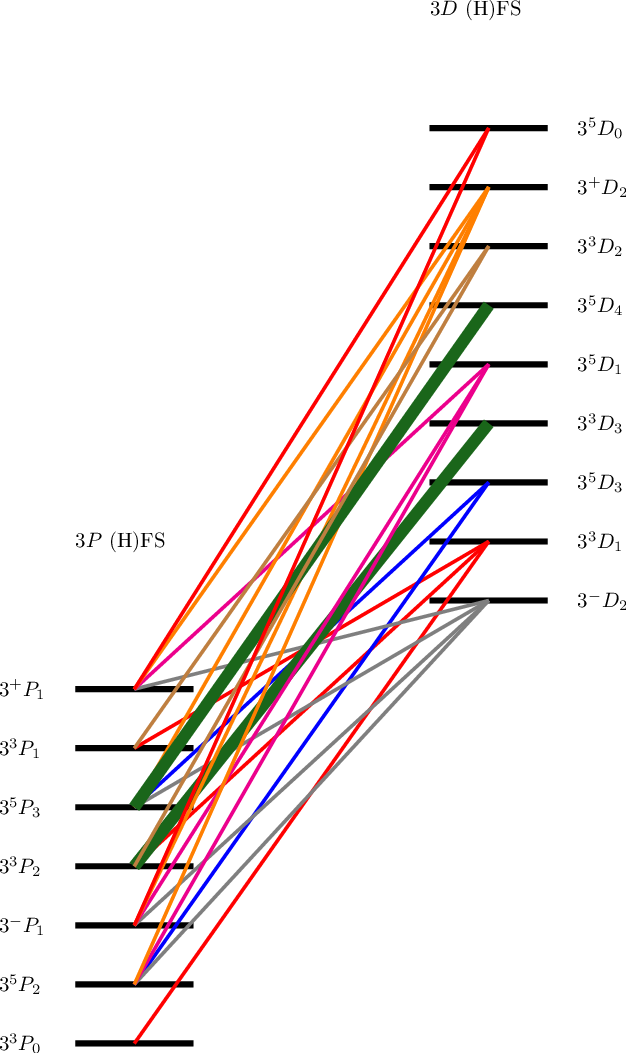}
\end{center}
\caption{\label{fig2} 
The dipole-allowed $3D$--$3P$ transitions are displayed graphically. 
The transitions are resolved with respect
to the (hyper)fine-structure of the levels [denoted as (H)FS].
When the upper sublevel undergoes only one 
dipole-permitted transition, we use
thick lines. Others transitions
are denoted by ordinary solid lines.
The viewgraph illustrates
the much more complex (hyper)fine-structure
that is encountered in the bound system of two
spin-1 particles, as compared to bound systems 
involving a light spin-1/2 particle orbiting a
heavy nucleus.}
\end{minipage}
\end{center}
\end{figure}

\begin{figure}[t!]
\begin{center}
\begin{minipage}{0.99\linewidth}
\begin{center}
\includegraphics[width=0.91\linewidth]{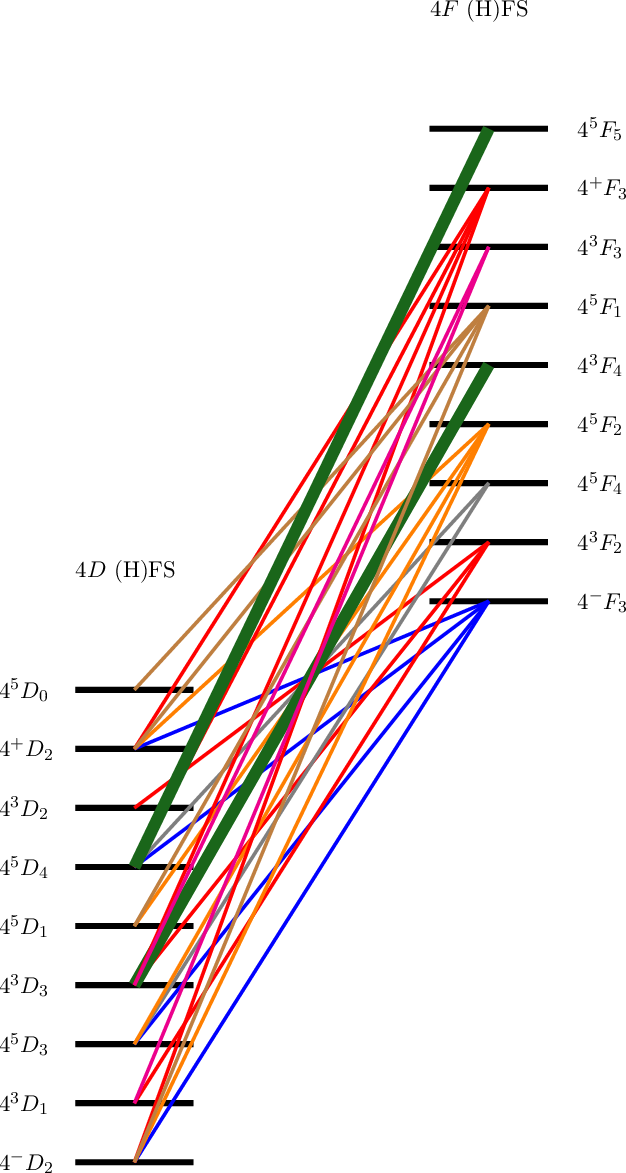}
\end{center}
\caption{\label{fig3} Same as the viewgraph 
Fig.~\ref{fig2},
but for the dipole-allowed $4F$--$4D$ transitions.}
\end{minipage}
\end{center}
\end{figure}

\begin{table}[t!]
\begin{minipage}{0.9\linewidth}
\caption{\label{table6} Energies and wavelengths for $3D \rightarrow 3P$
transitions in deuteronium.  The uncertainties for all of these transitions are
dominated by the strong interaction 
$3P$ state uncertainty of  $0.046\, {\rm eV}$.}
\begin{tabular}{c@{\hspace*{0.7cm}},@{\hspace*{0.91cm}},}
\hline
\hline\noalign{\smallskip}
transition 
  & \multicolumn{1}{c}{energy $\Delta E$}
  & \multicolumn{1}{c}{wavelength $\lambda$} \\
  & \multicolumn{1}{c}{${\rm (eV)}$} 
  & \multicolumn{1}{c}{${\rm (nm)}$} \\
\hline\noalign{\smallskip}
$3^-\!D_2 \rightarrow 3^-\!P_1$ & 1.x342(46) & 924x(32)  \\
$3^-\!D_2 \rightarrow 3^+\!P_1$ & 0.x936(46) & 1325x(65) \\
$3^-\!D_2 \rightarrow 3^5P_2$   & 1.x345(46) & 922x(32)  \\
$3^-\!D_2 \rightarrow 3^5P_3$   & 1.x134(46) & 1093x(45)  \\
\noalign{\smallskip}
$3^+\!D_2 \rightarrow 3^-\!P_1$ & 1.x410(46) & 879x(29)  \\
$3^+\!D_2 \rightarrow 3^+\!P_1$ & 1.x003(46) & 1236x(57) \\
$3^+\!D_2 \rightarrow 3^5P_2$   & 1.x412(46) & 878x(29)  \\
$3^+\!D_2 \rightarrow 3^5P_3$   & 1.x202(46) & 1032x(40) \\
\noalign{\smallskip}
$3^3D_1 \rightarrow 3^3P_0$ & 1.x452(46) & 854x(27)  \\ 
$3^3D_1 \rightarrow 3^3P_1$ & 1.x105(46) & 1123x(47)  \\
$3^3D_1 \rightarrow 3^3P_2$ & 1.x219(46) & 1017x(39)  \\
$3^3D_2 \rightarrow 3^3P_1$ & 1.x153(46) & 1075x(43)  \\
$3^3D_2 \rightarrow 3^3P_2$ & 1.x268(46) & 978x(36)  \\
$3^3D_3 \rightarrow 3^3P_2$ & 1.x245(46) & 996x(37)  \\
\noalign{\smallskip}
$3^5D_0 \rightarrow 3^-\!P_1$ & 1.x411(46) & 878x(29) \\ 
$3^5D_0 \rightarrow 3^+\!P_1$ & 1.x005(46) & 1234x(57) \\ 
$3^5D_1 \rightarrow 3^-\!P_1$ & 1.x410(46) & 879x(29)  \\ 
$3^5D_1 \rightarrow 3^+\!P_1$ & 1.x003(46) & 1236x(57) \\ 
$3^5D_1 \rightarrow 3^5P_2$   & 1.x412(46) & 878x(29)  \\ 
$3^5D_3 \rightarrow 3^5P_2$   & 1.x363(46) & 910x(31)  \\ 
$3^5D_3 \rightarrow 3^5P_3$   & 1.x152(46) & 1076x(43) \\ 
$3^5D_4 \rightarrow 3^5P_3$   & 1.x195(46) & 1038x(40) \\ 
\hline
\hline
\end{tabular}
\end{minipage}
\end{table}

\begin{table}[t!]
\begin{minipage}{0.9\linewidth}
\caption{\label{table7} Energies and wavelengths for $4F \rightarrow 4D$
transitions in deuteronium.  The uncertainties for these transitions arise from
the uncertainty in the quadrupole moment $Q_d$ and the strong interaction
uncertainty for 4D states (see Tables~\ref{table3} and~\ref{table5}).}
\begin{tabular}{c@{\hspace*{0.7cm}},@{\hspace*{0.91cm}};}
\hline
\hline\noalign{\smallskip}
transition 
  & \multicolumn{1}{c}{energy $\Delta E$} 
  & \multicolumn{1}{c}{wavelength $\lambda$} \\
  & \multicolumn{1}{c}{${\rm (meV)}$} 
  & \multicolumn{1}{c}{${\rm (nm)}$} \\
\hline\noalign{\smallskip}
$4^-\!F_3 \rightarrow 4^-\!D_2$ & 314.x9049(15) & 3937.x19(2)    \\
$4^-\!F_3 \rightarrow 4^+\!D_2$ & 286.x4151(14) & 4328.x83(2)    \\
$4^-\!F_3 \rightarrow 4^5D_3$   & 307.x2340(12) & 4035.x50(2)    \\
$4^-\!F_3 \rightarrow 4^5D_4$   & 289.x4061(10) & 4284.x091(15)  \\
\noalign{\smallskip}
$4^+\!F_3 \rightarrow 4^-\!D_2$ & 324.x9899(14) & 3815.x02(2)    \\
$4^+\!F_3 \rightarrow 4^+\!D_2$ & 296.x5000(14) & 4181.x59(2)    \\
$4^+\!F_3 \rightarrow 4^5D_3$   & 317.x3190(12) & 3907.x242(15)  \\
$4^+\!F_3 \rightarrow 4^5D_4$   & 299.x4910(10) & 4139.x830(14)  \\
\noalign{\smallskip}
$4^3F_2 \rightarrow 4^3D_1$     & 312.x2642(11) & 3970.x490(14)  \\ 
$4^3F_2 \rightarrow 4^3D_2$     & 291.x6724(11) & 4250.x80(2)    \\
$4^3F_2 \rightarrow 4^3D_3$     & 301.x3068(9)  & 4114.x882(12)  \\
$4^3F_3 \rightarrow 4^3D_2$     & 298.x6482(11) & 4151.x51(2)  \\
$4^3F_3 \rightarrow 4^3D_3$     & 308.x2826(9)  & 4021.x771(11)  \\
$4^3F_4 \rightarrow 4^3D_3$     & 304.x9837(8)  & 4065.x273(11)  \\
\noalign{\smallskip}
$4^5F_1 \rightarrow 4^5D_0$     & 295.x845(2)   & 4190.x86(3)  \\ 
$4^5F_1 \rightarrow 4^5D_1$     & 296.x5000(12) & 4181.x59(2)  \\ 
$4^5F_1 \rightarrow 4^-\!D_2$   & 324.x9899(15) & 3815.x02(2)  \\ 
$4^5F_1 \rightarrow 4^+\!D_2$   & 296.x5000(15) & 4181.x59(2)  \\ 
$4^5F_2 \rightarrow 4^5D_1$     & 292.x4153(11) & 4240.x00(2)   \\ 
$4^5F_2 \rightarrow 4^-\!D_2$   & 320.x9051(14) & 3863.x58(2)   \\ 
$4^5F_2 \rightarrow 4^+\!D_2$   & 292.x4153(14) & 4240.x00(2)   \\ 
$4^5F_2 \rightarrow 4^5D_3$     & 313.x2342(12) & 3958.x195(15) \\ 
$4^5F_4 \rightarrow 4^5D_3$     & 311.x1198(12) & 3985.x096(15)  \\ 
$4^5F_4 \rightarrow 4^5D_4$     & 293.x2918(9)  & 4227.x332(14)  \\ 
$4^5F_5 \rightarrow 4^5D_4$     & 299.x8126(9)  & 4135.x389(13) \\ 
\hline
\hline
\end{tabular}
\end{minipage}
\end{table}

%
%
\subsection{Selection Rules}
\label{sec6B}

One aspect we have not treated so far in this work concerns selection rules for
dipole transitions.  Here we work out the selection rules for electric dipole
transitions.  The interaction operator is proportional to $\vec r \cdot \vec
E$. The electric field acts to create a photon--our concern is with the
position operator $\vec r$, which involves the variables of the deuteronium
state.  Specifically, we find that single-photon electric dipole transitions
are allowed between states with quantum numbers $n_1,L_1,S_1,J_1,J_{z1}$ and
$n_2,L_2,S_2,J_2,J_{z2}$ (where $J_z$ represents the magnetic quantum number of the
total angular momentum) when
\begin{equation} \label{selection1}
\big \langle n_2,L_2,S_2,J_2, J_{z2} \big \vert (\vec r \, )_{1,m} 
  \big \vert n_1,L_1,S_1,J_1, J_{z1} \big \rangle \ne 0 \, .
\end{equation}
Here, $(\vec r\,)_{1,m}$ is the spherical tensor of rank one 
(and spherical component $m =-1,0,1$) 
corresponding to the position vector $\vec r$.  It is immediately clear
from the general rules of addition of angular momentum that a state with total
angular momentum $J_1$ and $z$ component $J_{z1}$, acted on by an operator with
angular momentum one and $z$ component $m$, only has overlap with a state of
total angular momentum $J_2=J_1 \pm 1$ or $J_2=J_1$, and $J_{z2}=J_{z1}+m$.  In
addition, parity conservation implies that $(-1)^{L_2}=(-1)^{L_1+1}$.
Additional conditions can be found by looking more deeply into the condition
\eqref{selection1}.  The states, as given in \eqref{psi}, have the form
\begin{equation}
\psi_{n LSJ M}(\vec r \, ) = R_{nL}(r) \, \Xi^{L S}_{J M}(\theta, \varphi)
\end{equation}
where
\begin{equation}
\Xi^{L S}_{J M}(\theta, \varphi)
= \sum_{M_{L},M_{S}} C^{J M}_{L M_L;S M_S} Y_{L M_L}(\theta, \varphi) \chi_{S M_S} \, ,
\end{equation}
where $\chi_{S M_S}$ is the spin-state of deuteronium, built from two spin-one
constitutents.  Also, one has $(\hat r)_{1m} = \sqrt{\frac{3}{4\pi}} Y_{1 m}
(\theta, \varphi)$. Hence, the angular part of Eq.~\eqref{selection1} is proportional to 
\begin{multline}
\sum_{M_{L_1},M_{L_2},M_{S_1},M_{S_2}} 
  C^{J_2 J_{z2}}_{L_2 M_{L_2};S_2 M_{S_2}}  
  C^{J_1 J_{z1}}_{L_1 M_{L_1};S_1 M_{S_1}} \\
\times \int \dd\Omega \, 
Y^*_{L_2 M_{L_2}}(\theta,\varphi) \,
Y_{1 m}(\theta,\varphi) \, Y_{L_1 M_{L_1}}(\theta,\varphi) 
\\
\times \langle \chi^\dagger_{S_2 M_{S_2}} \cdot \chi_{S_1 M_{S_1}} \rangle \,.
\end{multline}
Here, we denote the scalar product of the spin states
by a central dot ($\cdot$), for absolute clarity.
The spin inner product enforces the selection rule $S_2 = S_1$.  
The integral over the solid angle can be done for general 
quantum numbers [see Eq.~(6.64) of \cite{JeAd2022book}] and contains the $3J$ symbol 
\begin{equation}
\begin{pmatrix} L_2 & 1 & L_1 \\ 0 & 0 & 0 \end{pmatrix}
\end{equation}
as a factor, which vanishes unless $L_2 = L_1 \pm 1$. 
As a result, the selection rules for single-photon electric dipole transitions are
\begin{equation}
J_2=J_1 \pm 1 \; {\rm or} \; J_2=J_1 \, , \quad L_2 = L_1 \pm 1 \, , \quad S_2 = S_1 \, .
\end{equation}
The dipole allowed $3D$--$3P$ and $4F$--$4D$ transitions are displayed graphically in
Figs.~\ref{fig2} and \ref{fig3}.  Transition energies and wavelengths are shown
in Tables~\ref{table6} and \ref{table7}.

%
%
\section{Conclusions}

The bound system of a deuteron and its
antiparticle offers some unique 
elements in comparison with other two-body 
bound states where the partners have
equal mass. In comparison to positronium
\cite{BePe1980,Ka2004ijmpa,AdCaPR2022},
deuteronium offers drastically enhanced
characteristic field strengths, 
which allow the study of QED effects
under extreme conditions.
In comparison to protonium~\cite{Ba1989,KlBaMaRi2002},
deuteronium offers the added complexity of 
the spin-1 character of the bound particles,
which opens the door for an investigation
of the much more complex QED of 
spin-1 particles as compared to 
the well-understood spin-$1/2$ case.
In comparison to dimuonium (or, true muonium,
see Refs.~\cite{JeSoIvKa1997,KaIvJeSo1998,KaJeIvSo1998,BrLe2009}), 
deuteronium offers the enhanced stability 
of the deuteron, which does not
decay on a scale of microseconds (like the
muon and antimuon).
Hence, the deuteronium bound system
offers unique characteristics
for a detailed study of QED effects
of spin-1 particles in combination
with highly nontrivial spin-induced relativistic 
effects and radiative (vacuum polarization)
corrections.

Here, we have evaluated the dominant features of 
the spectrum of deuteronium. 
The main terms are the nonrelativistic Schr\"{o}dinger--Coulomb energy 
(order $\alpha^2 m_d$) and the one-loop eVP correction
(order $\alpha^3 m_d$).
The dominant spin dependence is contained in the 
Breit corrections (order $\alpha^4 m_d$, see Sec.~\ref{sec3}).
The muonic VP is enhanced in deuteronium (see Sec.~\ref{sec4B}),
and reducible as well as irreducible eVP corrections
are well under control (see Secs.~\ref{sec4C}---\ref{sec4E}).
In Sec.~\ref{sec5}, we have demonstrated that strong-interaction and
internal-structure effects do not represent obstacles
to a detailed study of the very interesting
spin-induced relativistic and radiative effects.
Namely, for non-$S$ states, the hadronic VP 
effect are numerically small (see Secs.~\ref{sec5A}
and~\ref{sec5B}).  Finite-size effects
can be treated perturbatively and are 
likewise numerically small for states with $L \ge 2$ (see
Sec.~\ref{sec5C}).  Strong-interaction corrections can be 
obtained on the basis of the generalized Deser-Trueman
formula~\cite{DeGoBaTh1954,BedH1955,By1957,Tr1961,KlBaMaRi2002,AdJe2025trueman}.

Certain Lamb shift 
transitions in deuteronium are in the optical or infrared range
and are easily accessible by laser spectroscopy.  
Two very attractive transitions appears to 
be the $3D$--$3P$ and $4F$--$4D$ Lamb shift transitions,
which are analyzed (with their dipole-allowed 
(hyper)fine-structure components in Figs.~\ref{fig2} and~\ref{fig3}. 
(See also Tables~\ref{table6} and \ref{table7}.)   
An additional important conclusion of our 
investigations is that internal-structure
corrections, for the mentioned transitions,
are numerically suppressed and do not
preclude precision experiments on the bound system
of a deuteron and its antiparticle.

\section*{Acknowledgments}

The authors acknowledge insightful discussions with 
Daniel Murtagh and Professor Jean-Marc~Richard.
This work was supported by the National Science Foundation through Grants
PHY-2308792 (G.S.A.) and PHY--2110294 (U.D.J.), and by the National Institute
of Standards and Technology Grant 60NANB23D230 (G.S.A.).

\appendix

\begin{widetext}

%
%
\section{Tensor Structures}
\label{appa}

We investigate some angular structures
relevant to the discussion. Let
$X^{(2) ij}$ and $S^{(2) ij}$ be two 
symmetric traceless second-rank Cartesian tensors,
with $i,j = x,y,z$,
\begin{equation}
X^{(2) ij} = X^{(2) ji}  \,,
\qquad
X^{(2) ii} = 0 \,,
\qquad
S^{(2) ij} = S^{(2) ji}  \,,
\qquad
S^{(2) ii} = 0 \,.
\end{equation}
The Einstein summation convention has been used.
The corresponding spherical tensors are denoted
as $X_{L m}$ and $S_{L m}$ with $L = 2$. 
We match the Cartesian against the spherical forms
by considering Eq.~(6.28c) of Ref.~\cite{JeAd2022book},
\begin{equation}
\label{norm_rank_2}
X_{2 0} 
= \sqrt{ \frac32 } \, X^{(2) 33} \,,
\qquad
S_{2 0}  
= \sqrt{ \frac32 } \, S^{(2) 33} \,.
\end{equation}
We denote Cartesian indices as superscripts, while
spherical tensor indices are denoted by subscripts.
The prefactor $\sqrt{3/2}$ follows from the use
of the Clebsch--Gordan coefficients, as explained
in Chap.~6 of Ref.~\cite{JeAd2022book}.
Of particular interest for us are
the second-rank tensors defined in terms of Cartesian 
vector components of the underlying vectors:
\begin{equation} 
X^{(2) ij} = (\hat x^i \, \hat x^j)^{(2)} = 
\hat x^i \, \hat x^j  - \tfrac13 \delta^{ij} \,,
\qquad
S_{a b}^{(2) ij} = ( S_a^i S_b^j )^{(2)} = 
\frac12 ( S_a^i S_b^j + S_a^j S_b^i ) - 
\tfrac13 \delta^{ij} \, \vec S_a \cdot \vec S_b \,.
\end{equation}
The latter is assumed to be the rank-two tensor 
constructed from $\vec\genspin_a$ and $\vec\genspin_b$.
The corresponding spherical tensors are normalized according to 
Eq.~\eqref{norm_rank_2},
\begin{equation}
X_{20} = \sqrt{\frac32} \, (\cos^2 \theta - \tfrac13) \,,
\qquad
(S_{ab})_{2 0} = \sqrt{\frac32} \,
( S_a^{3} \, S_b^{3} - \tfrac13 \vec S_a \cdot \vec S_b ) \,.
\end{equation}
One can show the following relation between the 
Cartesian and spherical formulations,
\begin{equation}
X^{(2) ij} \, S^{(2) ij}_{ab} = 
\sum_{q=-2}^2
(-1)^q \, X_{2q} \, (S_{a b})_{2,-q} \equiv \vec X_2 \cdot (\vec S_{ab})_2  \, ,
\end{equation}
which defines the scalar product of two rank-two spherical tensors 
[see Eq.~(5.2.4) of Ref.~\cite{Ed1957}]. The spherical tensors
$\vec X_2$ and $(\vec S_{a b})_2$ have spherical components 
$X_{2 q}$ and $(S_{a b})_{2 q}$ for $q \in \{ -2,-1,0,1,2 \}$.
The $\vec X_2$ tensor acts (multiplicatively) on orbital angular momentum
states.  The tensor $(\vec S_{a b})_2$ acts on spin-angular momenta states.
The total angular momentum is $\vec J$. Then, angular reduction according to
Eq.~(7.1.6) of Ref.~\cite{Ed1957} leads to 
\begin{equation} 
\label{input1}
\langle LS' J J_z | \vec X_2 \cdot (\vec S_{ab})_2 | LSJ J_z \rangle =
(-1)^{L + S' + J} \, 
\left\{ \begin{array}{ccc} J & S' & L \\ 2 & L & S \end{array} \right\} \,
\langle L || \vec X_2 || L \rangle \,
\langle S' || (\vec S_{ab})_2 || S \rangle \,.
\end{equation} 
One notes that, when $L = J$,  the $6J$ symbol
$\left\{ \begin{array}{ccc} J & S' & L \\ 2 & L & S \end{array} \right\}$
has a nonvanishing off-diagonal element in spin space
when $S'=0$ and $S = 2$, and when $S'=2$ and $S = 0$.
The angular tensor $\vec X_2$ is proportional to the $L=2$ spherical harmonic
according to $X_{2 q} = \sqrt{\frac{8 \pi}{15}} Y_{2 q}(\theta,\phi)$.  Its
reduced matrix element is
\begin{equation} 
\label{input2}
\langle L || \vec X_2 || L \rangle = 
\sqrt{\frac{2}{3}} (-1)^{L} (2L+1) 
\left( \begin{array}{ccc} L & 2 & L \\
0 & 0 & 0 \end{array} \right) \,.
\end{equation} 
The reduced matrix element for the single particle spin tensors (in states
built from two spin-1 particles) is
\begin{equation}
\label{input3}
\langle S' || (\vec S_{++})_2 || S \rangle = 
(-1)^{S} \, \sqrt{5} \, \Pi_{S' S}  
\left \{ \begin{array}{ccc} S & S' & 2 \\ 1 & 1 & 1 \end{array} \right \} \,,
\end{equation}
and
\begin{equation}
\label{input4}
\langle S' || (\vec S_{--})_2 || S \rangle = 
(-1)^{S'} \, \sqrt{5} \, \Pi_{S' S}  
\left \{ \begin{array}{ccc} S & S' & 2 \\ 1 & 1 & 1 \end{array} \right \} \,.
\end{equation}
We have, according to Eqs.~\eqref{input1},~\eqref{input2},~\eqref{input3} and~\eqref{input4},
\begin{eqnarray}
\label{D_LSpS J_appendix}
D_{L S' S J} &\equiv& 
\langle LS'J J_z | \, (\hat x^i \hat x^j)^{(2)} \, 
\big [ (S_+^i \, S_+^j)^{(2)} + (S_-^i \, S_-^j)^{(2)} \big ] 
| LSJ J_z \rangle 
\nonumber \\
&=& \langle LS'J J_z | \vec X_2 \cdot 
\big [ (\vec S_{++})_2 + (\vec S_{--})_2 \big ] | LSJ J_z \rangle 
\nonumber \\
&=& (-1)^{S'+J} \big [ (-1)^{S'} + (-1)^{S} \big ] \, 
(2L+1) \, \sqrt{ \frac{10}{3} } \, \Pi_{S' S} 
\left\{ \begin{array}{ccc} J & S' & L \\ 2 & L & S \end{array} \right\} \,
\left( \begin{array}{ccc} L & 2 & L \\ 0 & 0 & 0 \end{array} \right) \,
\left\{ \begin{array}{ccc} S & S' & 2 \\ 1 & 1 & 1 \end{array} \right\} \,.
\end{eqnarray}
The reduced matrix element of the mixed deuteron-antideuteron spin tensor can be written as
\begin{equation}
\label{input5}
\langle S' || (\vec S_{+-})_2 || S \rangle = 
(-1)^{(S'+S)/2} \left ( \frac{1+\delta_{S' S}}{2} \right ) \langle S' || \big [ (\vec S_{++})_2 + (\vec S_{--})_2 \big ] || S \rangle \, .
\end{equation}
It follows that
\begin{eqnarray}
\label{C_LSJ_appendix}
C_{LS'SJ} &\equiv& 
\langle LS'J J_z | \,  (\hat x^i \hat x^j)^{(2)} \, 
(S_+^i \, S_-^j)^{(2)} | LSJ J_z \rangle 
= \langle LS'J J_z | \vec X_2 \cdot (\vec S_{+-})_2 | LSJ J_z \rangle 
\nonumber \\
&=& (-1)^{(S'+S)/2} \left ( \frac{1+\delta_{S' S}}{2} \right ) D_{LS'SJ} \, .
\end{eqnarray}

\end{widetext}

\end{document}